\documentclass[10pt,journal,compsoc]{IEEEtran}

\ifCLASSINFOpdf
   \usepackage[pdftex]{graphicx}
   \DeclareGraphicsExtensions{.pdf,.eps,.png}
\else
\fi 

\usepackage[ngerman, english]{babel}
\usepackage[margin=1.2in]{geometry}

\usepackage{url} 
\usepackage{ctable}
\usepackage{multirow}
\usepackage{booktabs}

\usepackage[utf8]{inputenc}
\usepackage{threeparttable}
\usepackage{tabularx}
\usepackage{array}
\usepackage{algorithm}
\usepackage{algpseudocode}
 
\ifCLASSOPTIONcompsoc
\usepackage[caption=false,font=normalsize,labelfont=sf,textfont=sf]{subfig}
\else
\usepackage[caption=false,font=footnotesize]{subfig}
\fi

\ifCLASSOPTIONcompsoc
  \usepackage[nocompress]{cite}
\else
  \usepackage{cite}
\fi

\usepackage{color}
\usepackage{hyperref}

\usepackage{array}
\usepackage{mdwmath}
\usepackage{mdwtab} 
\usepackage{threeparttable}
\usepackage{multirow}
\usepackage{booktabs}
\usepackage{tabularx}
\usepackage{array}

\begin{document}

\setcounter{page}{1}
\setcounter{figure}{0} \renewcommand{\thefigure}{\arabic{figure}}

\title{Identification of Wearable Devices with Bluetooth}
\author{Hidayet~Aksu,
         A. Selcuk Uluagac,
~\IEEEmembership{Senior Member,~IEEE,}
        and~Elizabeth~S.~Bentley
\IEEEcompsocitemizethanks{\IEEEcompsocthanksitem H.~Aksu and A.~S.~Uluagac are with the Dept. of Electrical and Computer Engineering, Florida International University, Miami,
FL, 33174. \protect\\
E-mail:\{haksu,suluagac\}@fiu.edu.
\IEEEcompsocthanksitem E.~S.~Bentley is with Air Force Research Lab., Rome, NY 13441-4514. \protect\\ 
E-mail: elizabeth.bentley.3@us.af.mil }
\thanks{(©2018 IEEE)}}

\markboth{IEEE Transactions on Sustainable Computing,~Vol.~X, No.~X, Month~Year}%
{Aksu \MakeLowercase{\textit{et al.}}: \title }

\IEEEtitleabstractindextext{%
\begin{abstract}
 With wearable devices such as smartwatches on the rise in the consumer electronics market, securing these wearables is vital. However, the current security mechanisms only focus on validating the user not the device itself. Indeed, wearables can be (1) unauthorized wearable devices with correct credentials accessing valuable systems and networks, (2) passive insiders or outsider wearable devices, or (3) information-leaking wearables devices. Fingerprinting via machine learning can provide necessary cyber threat intelligence to address all these cyber attacks. In this work, we introduce a wearable fingerprinting technique focusing on Bluetooth classic protocol, which is a common protocol used by the wearables and other IoT devices. Specifically, we propose a non-intrusive wearable device identification framework which utilizes 20 different Machine Learning (ML) algorithms in the training phase of the classification process and selects the best performing algorithm for the testing phase. Furthermore, we evaluate the performance of proposed wearable fingerprinting technique on real wearable devices, including various off-the-shelf smartwatches. Our evaluation demonstrates the feasibility of the proposed technique to provide reliable cyber threat intelligence. Specifically, our detailed accuracy results show on average 98.5\%, 98.3\% precision and recall for identifying wearables using the Bluetooth classic protocol.
\end{abstract}
 }

 \onecolumn
 
 \setcounter{page}{0}

 This work has been accepted in IEEE Transactions on Sustainable Computing.

\href{https://doi.org/10.1109/TSUSC.2018.2808455}{DOI: 10.1109/TSUSC.2018.2808455}

URL: \url{http://ieeexplore.ieee.org/stamp/stamp.jsp?tp=&arnumber=8299447&isnumber=7742329}\\
\\\\

IEEE Copyright Notice:\\
© 2018 IEEE. Personal use of this material is permitted. Permission from IEEE must be obtained for all other uses, in any current
or future media, including reprinting/republishing this material for advertising or promotional purposes, creating new collective
works, for resale or redistribution to servers or lists, or reuse of any copyrighted component of this work in other works. 

 \newpage
\maketitle
\IEEEdisplaynontitleabstractindextext
\IEEEpeerreviewmaketitle 
 
\IEEEraisesectionheading{\section{Introduction}\label{sec:Intro}}
\IEEEPARstart{C}{yberspace}
is expanding rapidly with the introduction of new Internet of Things (IoT) devices. Today, it is extremely challenging to find a device without any Internet connection capability. Wearables, smart watches, glasses, fitness trackers, medical devices, and Internet-connected house appliances have grown exponentially in a short period of time. It is estimated that on average, one device is assumed to be connected to Internet today every eighty seconds and our everyday lives will be dominated by billions of smart connected devices by the end of this decade~\cite{web:Kelly}. Indeed, it is predicted that by 2020, there will be 50 to 100 billion devices connected to the Internet~\cite{IoT_Evolution,IoT_Europe}, forming a massive IoT. This emerging IoT technology will drastically change our daily lives and enable smarter cities, health, transportation, and energy~\cite{IoT_Europe}. Among these devices, a considerable number of them will be the wearable devices that can be carried by individuals such as watches, fitness bands, sensors (e.g., heart-rate, stride), etc. By 2019, it is estimated that one in four smartphone owners will also be using a wearable device~\cite{watch}.

On the other hand, one of the relatively overlooked problems in the industry or in any networking environment today is that although wearable device vendors follow the general guidelines while implementing a specific software, hardware, or firmware to be compatible with the industry standards and other technologies, they, unfortunately, do not fully comply with the specifics of the standards~\cite{miller2005coping, caballero2007fig}. Different implementations of the same functionality can be observed with different vendors due to differences in interpretations and lenient parts of the standards. Similarly, it is possible that counterfeit wearable devices or devices with corrupted hardware or software components may exist in a networked environment without the knowledge of the network administrator~\cite{DODJan13Report, web:Stecklow}. These wearable devices may participate in the regular data collection transactions, glean important information from unwary benign devices nearby, and leak such information to adversaries~\cite{DODJan13Report, web:Stecklow}.  Moreover, a  network can dynamically grow and shrink in size with new wearable devices and equipment depending on the needs. For instance, employees can bring their own wearables (aka BYOD) to their networks. New wearable devices can join and leave an authorization realm and device configurations can change dynamically, or even more frequently than other traditional networking settings. Similarly,   
a wearable device can be compromised or can be made ineffective by adversaries or, simply, a small wearable   
device can be dropped or lost in a networking environment. In such a case, a wearable device could still be part of an authentication realm acting as an insider threat to the other legitimate operational networked resources. More specifically, new devices can be (1) unauthorized wearable devices with correct network credentials, (2) passive insiders or outsider wearable devices, or (3) information-leaking wearable devices. 

Furthermore, most of the wearables are resource-limited and have limited processing capabilities. This poses challenges to the most state-of-the-art security solutions. For instance, an insider attack could be avoided with a multi-factor authentication mechanism. However, achieving a multi-factor authentication~\cite{Wenyi-Maca-InfocomWS-14} on a wearable device with limited resources may be challenging, if not impossible.  \textit{To alleviate these concerns, in this work, we introduce machine learning based wearable fingerprinting tool as a non-intrusive complementary security mechanism for wearables}. Such a fingerprinting mechanism does not solely depend on current security solutions to verify whether the device whose security is questioned is actually the device it claims to be but incorporates obtained cyber threat intelligence.  
 With fingerprinting, unauthorized wearable devices via their reproducible fingerprints, possibly inserted by authorized individuals (aka insiders) can be detected. Furthermore, a wearable fingerprinting mechanism can also be utilized to identify unmanageable wearables without any sophisticated software architecture on them. Hence, a wearable fingerprinting mechanism can supplement current security solutions (e.g., access control and authentication) to gain more information and confidence in critical decisions when classical security solutions cannot be efficiently operated in a wearable realm. 
  For instance, a Bluetooth speaker may be needed in a conference room. Although this speaker will be connected to the network, traditional credential  or NAC-based solutions are not applicable as the speaker does not support such services. However, network admins may want to make sure any rogue device with Bluetooth support cannot access the network. Network access is granted to a device only after it is identified as an expected device.

 In this work, we propose to utilize the timing information of Bluetooth classic protocol. This protocol is predominately used by the wearable devices in the market today. Our framework utilize a comprehensive set of Machine Learning (ML) algorithms (20 different ML algorithms) in training phase of the classification process to pick the best performing algorithm. \textit{To the best of our knowledge, the proposed fingerprinting technique, as well as the intelligent utilization of 20 different ML algorithms, is the first in the wearables realm}. Moreover, we apply our wearable fingerprinting technique on different wearables, i.e., various smart watches. \textit{Our detailed evaluation demonstrates the functionality and feasibility of the proposed technique with 98.5\%, 98.3\% precision and recall for wearables using Bluetooth classic protocol}.  
 
The remainder of this manuscript is organized as follows. We review the related work in Section~\ref{sec:relatedWork}. Background on wearables is presented in Section~\ref{sec:background}. We also explain different fingerprinting techniques in the same section. In ~\ref{sec:SystemModel}, we introduce the components of our wearable fingerprinting framework.  
In Section~\ref{sec:empiricalanalysis}, we discuss test setup and provide the empirical analysis of wearables using Bluetooth classic protocol. Then in Section~\ref{sec:securityImpact}, we discuss how fingerprinting can be used to complement security and how a real operational wearable networking environment could benefit from fingerprinting. Finally, we conclude this paper in Section~\ref{sec:Conclusion}.

\section{Related Work}\label{sec:relatedWork} 
There is currently no work that fingerprints wearable devices. However, fingerprinting has been applied by some many earlier studies.   
A seminal work in this area was introduced by Kohno et al. in~\cite{1059392}. In~\cite{1059392}, a method for remotely fingerprinting a physical device by exploiting the implementation of the TCP protocol stack was proposed. The authors use the TCP timestamp option of outgoing TCP packets to reveal information about the sender's internal clock. The authors' technique exploits microscopic deviations in the clock skews to derive a clock cycle pattern as the identity for a device. The authors of~\cite{1409958} take a similar approach to that of~\cite{1059392} (i.e., using clock skews to uniquely identify nodes), however, the goal of~\cite{1409958} is to uniquely fingerprint access points (APs), obtaining the timestamp from 802.11 beacon frames. Similarly, the authors in \cite{WDF:5a} use timing information between commands and responses on the Universal Serial Bus (USB) to distinguish between variations in model identifiers, OSs (and sometimes OS version number), and whether a machine is answering from a real or virtual environment.  
There have also been other physical layer approaches to fingerprint wireless devices. A good survey on the physical-layer identification of wireless devices can be found in~\cite{Danev:2012:PIW:2379776.2379782}. Radio frequency (RF) emitter fingerprinting uses the distinct electromagnetic (EM) characteristics that arise from differences in circuit topology and manufacturing tolerances.  This approach has a history of use in cellular systems and has more recently been applied to Bluetooth~\cite{bluetooth} and Wi-Fi~\cite{wifi} emitters. The EM properties fingerprint the unique transmitter of a signal and differ from emitter to emitter. This technique requires expensive signal analyzer hardware to be within RF range of the target. In a more recent work~\cite{Uluagac-GTID-CNS,Sakthi-GTID-TDSC}, the authors developed a passive wired-side technique to fingerprint types of devices connected to a Wireless Local Area Network (IEEE 802.11 g/n). Different from this WLAN fingerprinting work, our identification focuses on the characteristics of Bluetooth classic protocol, which is mostly used by the resource-limited wearables. Also, their work considers one type of classification mechanism (i.e., artificial neural networks) whereas our fingerprinting framework is able to utilize 20 different Machine Learning algorithms to determine the best performing one for fingerprinting problem. Finally, a recent useful survey of fingerprinting mechanisms can be found in~\cite{7239531}.

\section{Background }\label{sec:background} 
\subsection{Wearables}
It is possible to see the early examples of wearables  mostly in the smart watch realm. For smart watches, there are four major smart watch operation system vendors: Android-based, IOS-based, Samsung-based with Tizen O/S~\cite{web:Tizen}, and Pebble-based with Pebble O/S~\cite{web:pebble}.  The Pebble O/S is based on an open source Real Time Operating System (FreeRTOS) for embedded devices while the Tizen is another open-source Linux-derivative operating system.  
In this paper, we only focus on Android-based ones given their popularity with their open-source nature.  
 \begin{figure}[t!]
  \centering
\includegraphics[width=0.40\textwidth]{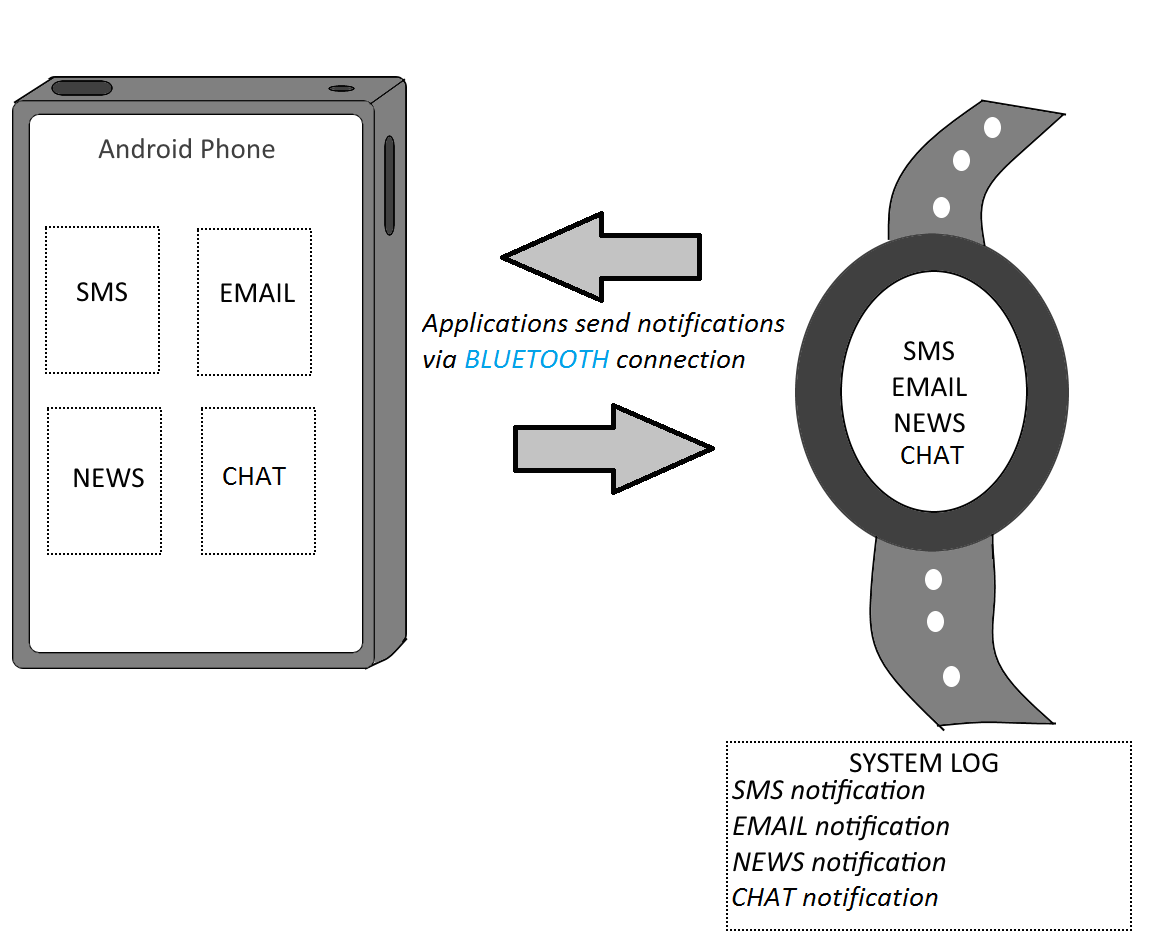}
  \caption{Overall Sync Architecture between the Phone and Wearable. }
  \label{fig:system}
\end{figure}

A wearable device (e.g., smart watch, fitness band)  
usually needs to work and synchronize with another more resourceful Android device such as a tablet or smartphone to be fully functional (Figure~\ref{fig:system}). The Android Wear app is the primary conduit for communication between an Android Wear device and a smartphone/tablet. Without the application installed and running on the Android handheld device, the Android Wear and the handheld are unable to pair, limiting the serviceability of the Android Wear technology. Moreover, the operating system of the Android handheld must be running on Android 4.3 Jelly Bean or higher.

\begin{figure}[t!]
    \centering
    \includegraphics[width=0.40\textwidth]{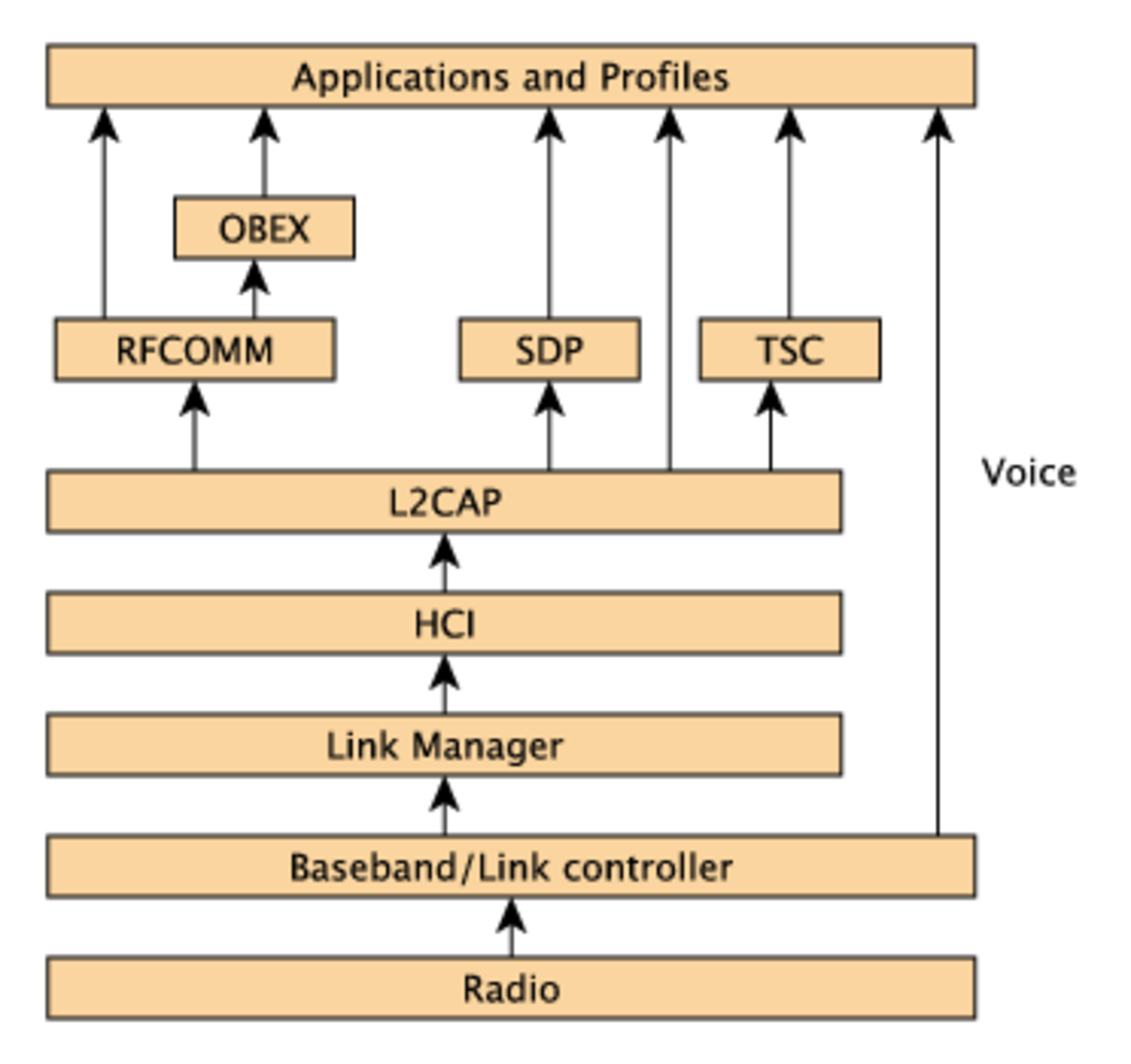}
    \caption{Bluetooth classic protocol stack}
    \label{fig:bluetooth-classic-protocol-stack}
\end{figure}
For most of the wearable devices, the communication occurs via Bluetooth  
protocols.  
An example communication between an Android wear smart watch and a smartphone is illustrated in Figure~\ref{fig:system}. In this example, the smartwatch and the smartphone use the \textit{notifications} over Bluetooth. With notifications, the wear devices and applications share information among themselves. The sending application for the notifications can be on one device and the receiving application on another. When a notification is created on a smartphone, it is sent to the Android Wear application, which then sends the notification to the synced wearable. The overall architecture is shown in Figure~\ref{fig:system}. Notifications from all applications on the smart device are sent to the wearable via the Android Wear application using a Bluetooth  
connection. These notifications are immediately displayed on the wearable's screen.

\subsection{Wearables with Bluetooth Classic}
Bluetooth classic is the legacy version of Bluetooth, which is first created in 1994~\cite{web:bluetoothdate} while the Bluetooth Special Interest Group (SIG) is formed in 1998~\cite{web:bluetoothsig}. It is widely used in the market today and it is also known as Bluetooth BR/EDR (basic rate/enhanced data rate). The current version is Bluetooth v4.2, which was released in December 2014. Figure~\ref{fig:bluetooth-classic-protocol-stack} displays Bluetooth classic protocol stack. 

Meanwhile, Bluetooth security depends on pairing process and use of authentication and encryption. All security features depend on device name, address, i.e., BDADDR, and encryption keys. Device name and address can be spoofed while encryption keys can be copied to other devices. Thus, enabling device fingerprinting, which depends on device hardware, can increase the overall security level of Bluetooth speaking wearables.

\subsection{Fingerprinting Wearables}\label{sub-sec:fingTypes}  
Wearable fingerprinting can generally be achieved in at least two different ways.  
The first depends on the goal of fingerprinting and there can be two types of fingerprinting under this: \textit{Device} or \textit{device type fingerprinting}. With the \textit{device fingerprinting}, individual devices are fingerprinted. The main goal is to distinguish an individual device from another one of its kind. For instance, assume there are two smart watches from the same vendor, Sony Smart watch 1 and Sony Smart watch 2, the goal is to distinguish one from the other. With the \textit{device type fingerprinting}, devices from different vendors are identified. The identification is based on the vendor diversification rather than the individual devices by the same vendors.     
\begin{figure}[t!]
  \centering
  \includegraphics[width=.99\linewidth]{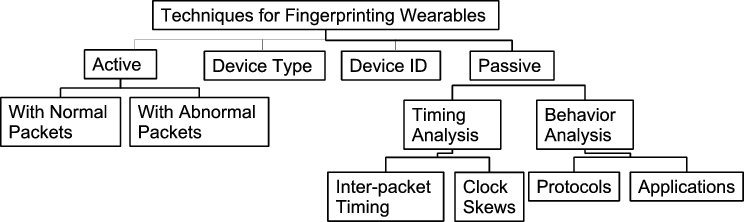}
  \caption{Fingerprinting wearables}
  \label{fig:WearableFingerprinting}
\end{figure} 

The Second type of fingerprinting depends on the method of fingerprinting. In this category, there can be two generic ways: 
 \textit{active} and \textit{passive fingerprinting}~\cite{5686874}. 
 
Under the \textit{active fingerprinting} technique, the wearable devices are fingerprinted with an external stimulus (e.g., a specific packet) and the results returned from the device are fingerprinted and analyzed for signature generation and identification, respectively. Specifically, the active fingerprinting technique involves two complementary approaches. In the first one, \textit{regular packets} are sent to the wearable devices whereas, in the second one, \textit{malformed packets} (i.e., abnormal, unexpected) are sent to the wearables. Regular packets sent to the devices conform to the general rules of the standards and protocols.  Responses coming from the wearable devices, the difference in content, length, orders of packets, packet arrival, inter-arrival time can all be observed and analyzed in the fingerprinting process and be part of the signature generation process. 

In the \textit{passive fingerprinting} technique, the wearable devices are monitored for the information they carry (e.g., protocol packet fields) or generate (e.g., timing analysis between packets) and this observed information is the basis for the signature generation and the identification afterward. The passive fingerprinting can be accomplished via two different approaches. In the first one, \textit{behavior analysis}, the protocols, the applications, the protocol headers, protocol fields that are sent in the clear are all observed for how they behave and how differences in implementations of protocols and applications vary across devices. It is widely known that each vendor implements enhanced versions of certain functions although they conform to the industry standards. So, diversity in protocol functionalities can be observed. For instance, Bluetooth stack may have been implemented using different versions of the Bluetooth protocol. With this type of fingerprinting, it is possible to catch device types easier than individual devices within a specific device class. 
	
The second passive fingerprinting approach involves observing the timing patterns (e.g., interpacket-arrival times (IAT)) between the communicating end-points (e.g., wearable-to-wearable or wearable-to-other smart equipment communication)~\cite{Uluagac-GTID-CNS,Sakthi-GTID-TDSC}. The third one involves the observation of the clock skews~\cite{1059392}. Estimated clock skews from the fingerprinter's point of view can provide a good opportunity for generating unique device signature as each wearable device vendor may have differing internal clocks.  It is important to note that the last two approaches are non-intrusive, does not require deep-packet inspection,  and can be applied to any type of traffic whether the traffic is encrypted or not. 
 
In both active and passive methods, versatile open source tools such as Python's Scapy~\cite{web:Scapy} can be utilized for generating the regular or abnormal packets and open source software-based (e.g., Tcpdump~\cite{web:tcpdump}, Wireshark~\cite{web:wireshark}) or hardware-based (e.g., Ubertooth~\cite{web:ubertooth}) packet sniffers can all be used for monitoring and capturing the wearable traffic. The captured results are analyzed after the process or in-realtime.

Moreover, the aforementioned active and the passive wearable fingerprinting techniques can be applied to two different components of the wearable device: Applications and protocols. These important components interact with the outside world and provide an excellent opportunity for fingerprinting. For instance, requesting a special information (e.g., nonexistent) from a wearable device via the active fingerprinting method and observing its response may provide different results depending on the application type on the wearable device. Further, a certain application (e.g., a notification mechanism) or a protocol over different-but-seemingly similar wearable devices (e.g., Sony Smart watch vs. LG Smart watch) may generate different observable signatures.  The different fingerprinting methodologies that can be utilized in a wearable fingerprinting framework are summarized in Figure~\ref{fig:WearableFingerprinting}. Note that \textit{in this work, we utilize an inter-packet timing-based timing analysis method on Bluetooth classic protocol packets from wearable devices, effectively combining two different techniques of passive fingerprinting, timing and behaviour analysis to determine the device type}. The details of our fingerprinting framework are given in the following section.

\section{Wearable Fingerprinting Framework }\label{sec:SystemModel} 
\begin{figure}[tb]
  \centering
  \includegraphics[width=.95\linewidth]{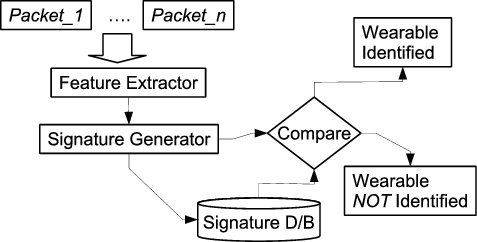}
  \caption{Wearable fingerprinting framework.}
  \label{fig:FingerprintingFramework}
  \vspace{-0.5cm}
\end{figure}

Our wearable fingerprinting framework consists of four main components as shown in Figure~\ref{fig:FingerprintingFramework}. (1) Packet capture; (2) Feature extractor; (3) Signature generator; and (4) Comparison. In this section, we articulate these briefly. 

\begin{itemize}
\item \textit{Packet capture}: The first step in our wearable fingerprinting framework involves capturing Bluetooth classic packets 
from a wearable device. Note that Bluetooth classic is predominantly used in the wearables domain. 

\item \textit{Feature extractor}: As the packets are collected from the wearable devices, this component is responsible for extracting the features from the packets. In our framework, distinguishing information is inter-arrival-times between Bluetooth packets. 
\item \textit{Signature generator}: The next component in the fingerprinting is to generate the signatures using the features as the basis to reveal patterns in the data. In our work, the signatures are probability distributions.    
Once the signatures are generated, they are also stored in a database so that they can be used to compare with new signatures when identifying unknown wearables. In other words, signatures are used to train the prediction models for various ML algorithms. 
\item \textit{Comparison/Prediction}: The final step in the wearable fingerprinting process involves comparing a stored signature with the wearable that  needs to be identified. If a match with a known signature is found for the unknown wearable, the unknown wearable is deemed identified, otherwise unidentified. It should be noted that from a security stand, both results are valuable.  
Here trained machine learning models are used to make predictions.

\end{itemize}

\begin{algorithm}[htb]
\centering
   \scriptsize
     \caption{Pick Best ML Algorithm - Training Phase } \label{alg:pickbest}
\begin{tabbing} 
    {\bf\ Input}: \hspace{0.3cm}\= $ds$: learning dataset, 
      \\\hspace{1.40cm} $algs$: list of supported ML algorithms,
     \\\hspace{1.35cm} $filters$: list of filters  \\
{\bf\ Output}: returns best performing algorithm\>
\end{tabbing}
\begin{algorithmic}[1] 
	 \State $v \gets $ empty vector
     \For{each algorithm $alg$ in $algs$ list }
         \For{each filter $f$ in $filters$ list } 
		    \State $ds_f \gets apply~filter(ds,f)$ 
		    \State $feature\_set \gets generate~signature(ds_f)$ 
		    \State $model_{alg} \gets build~model(feature\_set, alg)$ 
		    \State $accuracy_{f,alg} \gets test~model(model_{alg},ds_f)$ 
		    \State add pair $<alg, accuracy_{f,alg}>$ to vector $v$
		\EndFor
	\EndFor
	\State $v_{top15} \gets filter~top~15~percent~by~accuracy(v)$
	\State $alg\_freq \gets compute~frequency~per~algorithm(v_{top15})$
	\State $alg \gets most~frequent~algorithm(alg\_freq)$
	\State \Return $alg$
 	\end{algorithmic}
 \end{algorithm} 

Our wearable fingerprinting framework makes use of classifiers from Weka~\cite{Hall:2009:WDM:1656274.1656278} project and also an external neural network implementation~\cite{web:WekaNeuralNetwork}. 
The ML algorithms we used are listed in Table~\ref{tab:classifiers}. Further, this table includes the type of the algorithms. As different ML algorithms can model different patterns in data, we know that the best performing algorithm will depend on the pattern inside Bluetooth traffic captures and extracted features. However, there is no prior knowledge about the true nature of the pattern in this data. So, we selected well-known algorithms designed to model various patterns. For instance, LibSVM and SimpleLogistic like algorithms capture functional patterns while BayesNet and NaiveBayes like algorithms capture stochastic patterns. Thus, we incorporated algorithms from all major pattern types as listed in the Table~\ref{tab:classifiers} and exhaustively used them.
Our framework is able to pick the best performing ML algorithm among all the supported ones in the training phase using the Algorithm~\ref{alg:pickbest}. Specifically, the algorithm \textit{Pick Best ML Algorithm} uses the training dataset $ds$, list of supported ML algorithms $algs$, and the list of filters $filters$. It computes accuracy for each ML algorithm $alg$ and each filtering on training data and keeps a vector of $<alg,accuracy>$ pairs. Then, it selects top 15 percent of best performing algorithms and compute the frequency of each algorithm in this top list. The most frequent algorithm is picked as the best algorithm. Hereafter, this algorithm is used in the testing phase.

Then, dataset to signature generation is shown in Algorithm~\ref{alg:gensig}. In order to generate the signature, first,  inter arrival time vector $iat$ is computed from the input dataset. Then, the density distribution of $iat$ vector is generated. Finally, this distribution is converted into a histogram and each bin height in the histogram becomes a feature in the signature. 

\begin{table}[t]
\scriptsize
  \centering
    \caption{ML Classifiers used by our wearable identification  framework.} 
  \begin{tabular}{  l  p{6cm} }
   \textbf{Type} & \textbf{Name} \\ \hline\hline
   Functions & LibSVM, MultilayerPerceptron, NeuralNetwork, SMO, SimpleLogistic  \\ \hline
  Bayes & BayesNet, NaiveBayes, NaiveBayesMultinomialUpdateable, NaiveBayesUpdateable  \\ \hline
  Rules & DecisionTable, JRip, OneR, PART \\ \hline
  Trees & DecisionStump, HoeffdingTree, J48, LMT, REPTree, RandomForest, RandomTree  \\ \hline
  \end{tabular}
    \label{tab:classifiers}
\end{table}

\begin{algorithm}[H]
\scriptsize
     \caption{Generate~Signature - All Phases } \label{alg:gensig}
\begin{tabbing} 
    {\bf\ Input}: \hspace{0.3cm}\= $ds$: dataset   \\
{\bf\ Output}: returns $ds$ signature\>
\end{tabbing}
\begin{algorithmic}[1] 
	 \State $iat \gets extract~inter~arrival~time(ds)$
	 \State $dd \gets generate~density~distribution(iat)$
	 \State $features \gets converts~to~features(dd )$
 	\State \Return $features$
 	\end{algorithmic}
 \end{algorithm}

\begin{table}[htb]
 \scriptsize
  \centering
    \caption{Paired smartphone and wearable devices in our tests using Bluetooth BR/EDR protocol.}\label{table:bluetoothdevices}
    \begin{tabular}{  l  l  l  l } 
\textbf{Device Type} & \textbf{Make} & \textbf{Marketing Name} & \textbf{OS} \\ \hline \hline
Smartwatch & Sony & Sony smart watch & Android \\
Smartphone & Samsung & Galaxy S5 & Android \\
Smartwatch & Motorola & Moto 360 & Android \\
Smartwatch & Asus & ZenWatch  & Android \\
Smartwatch & LG & G Watch R  & Android \\
Smartwatch & LG & LG Urbane  & Android \\
Smartwatch & Samsung & Gear Live & Android \\
    \hline
    \end{tabular}
\end{table}

\section{Performance Evaluation} \label{sec:empiricalanalysis}
 In order to evaluate the feasibility and efficacy of the proposed wearable  
fingerprinting framework, we setup a testbed with a set of representative wearable devices and studied different test scenarios empirically.

\subsection{Testbed and Experiment Methodology}
\begin{figure}[htb]
    \centering
    \includegraphics[width=0.45\textwidth]{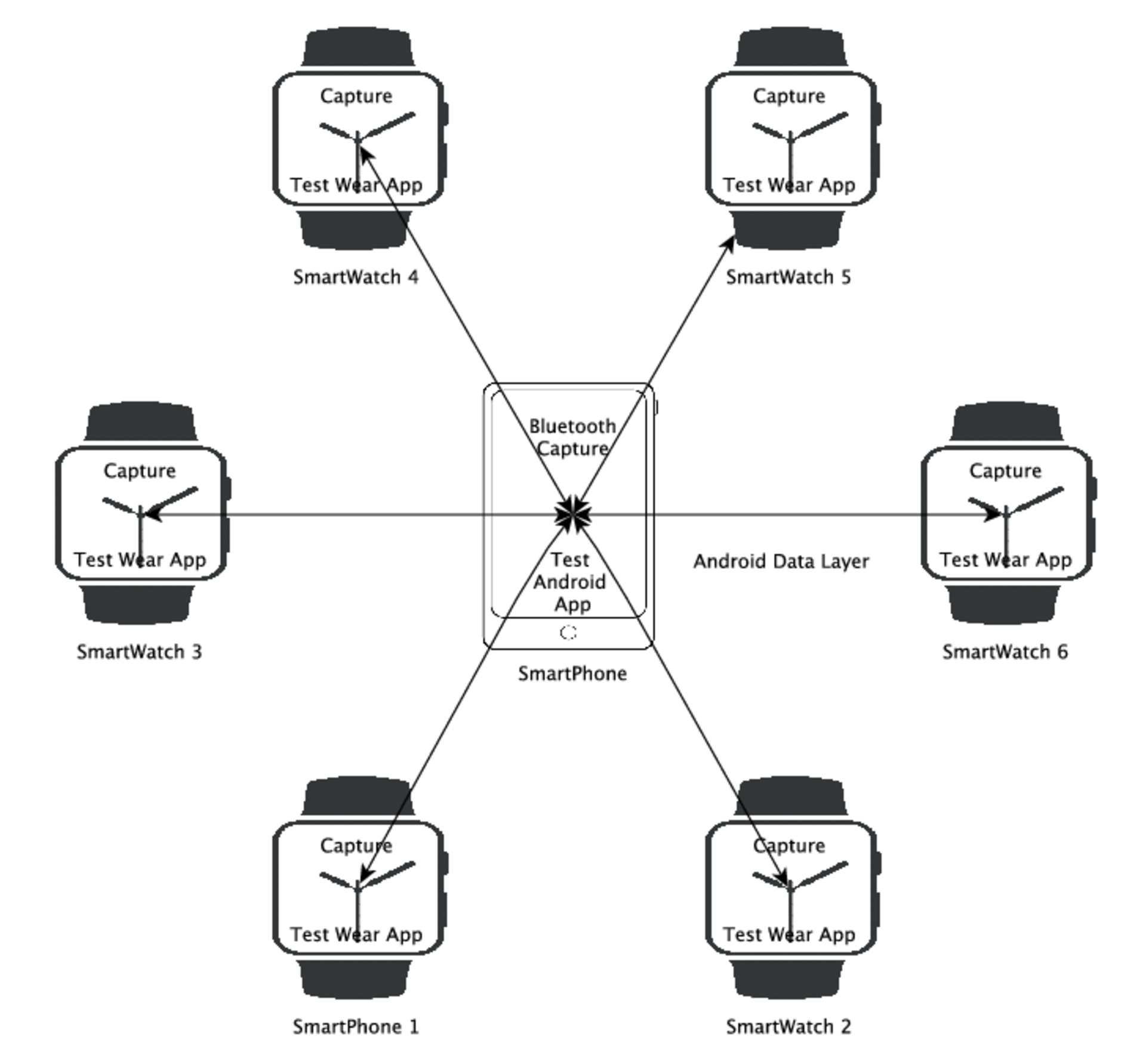}
    \caption{Bluetooth classic testbed with wearables}
    \label{fig:classictestbed}
\end{figure}

We setup a testbed of six smart watches and a master smartphone.  Table~\ref{table:bluetoothdevices} provides the list of Bluetooth classic wearable devices we used in our tests. 
All watches were paired with the master smartphone via the Android Wear app as illustrated in Figure~\ref{fig:classictestbed}. Note that the smart watches need this Android Wear App running on the paired smartphone.  
In this way, a communication between the smartphone and smart watch is established as explained in Section~\ref{sec:background}. 
Android Wear App pushes some limitations on Bluetooth connectivity such as only smartphone-to-smart watch communication is supported. Although a smartphone can be paired with many smart watches, a smart watch can be paired with only one smartphone at a time.  Since we want to profile usual Bluetooth traffic on Wearables, we followed the limitations associated Android Wear. All Bluetooth communication was captured at the smartphone.   

Smart watches receive all notifications from the paired smartphone. Also, any app in the smartphone  communicates with its wearable extension over the Bluetooth channel. This communication is managed by Android APIs. To facilitate this, we developed a test Android app with its wearable extension. This test app generated the Bluetooth traffic between the smartphone and smart watch.  
The generated traffic contained random-sized notifications from the smart watch to the smartphone and from phone to watch. Each experiment included about 300 exchanged messages. For each smart watch, we repeated the experiment 40 times. The total amount of captured Bluetooth traffic was about 1.08 GBytes.

\subsection{Experiments and Results} \label{sec:classicexpresults}
As articulated earlier, our evaluation focused on the packet inter-arrival-time (IAT) distributions as the fundamental wearable fingerprinting feature. In order to determine the best filtering cases for IATs, we analyzed the packet captures with \textit{packet length} and \textit{Bluetooth protocol} breakdowns.

\begin{table*}[htbp]
    \caption{Hardware specs of tested wearables and a smartphone with Bluetooth classic}
    \label{table:classicdevicedetails}
  \centering
   \scriptsize
    \begin{tabular}{ l l l l l} 
    
    \textbf{Device} & \textbf{Chipset} & \textbf{CPU} & \textbf{RAM}  & \textbf{Bluetooth} \\ \hline\hline
Sony smartwatch & Qualcomm Snapdragon 400 APQ8026 & ARM Cortex-A7, 1200 MHz, 4 Core & 512 MB  & 4.0 \\\hline
Galaxy S5 & Qualcomm Snapdragon 801 MSM8974AC & Krait 400, 2.5 GHz, 4 Core & 2 GB LPDDR3 & 4.0 \\\hline
Moto 360 E9A4 & Texas Instruments OMAP 3 3630 & ARM Cortex-A8, 1.2 GHz, 1 Core & 512 MB LPDDR & 4.0 \\\hline
ZenWatch 1726 & Qualcomm Snapdragon 400 APQ8026 & ARM Cortex-A7, 1.2 GHz, 4 Core & 512 MB  & 4.0 \\\hline
G Watch R 4050 & Qualcomm Snapdragon 400 APQ8026 & ARM Cortex-A7, 1.2 GHz, 4 Core & 512 MB  & 4.0 \\\hline
LG Urbane CFA0 & Qualcomm Snapdragon 400 APQ8026 & ARM Cortex-A7, 1.2 GHz, 4 Core & 512 MB  & 4.1 \\\hline
Gear Live 3103 & Qualcomm Snapdragon 400 APQ8026 & ARM Cortex-A7, 1.2 GHz, 4 Core & 512 MB LPDDR2 & 4.0 \\\hline
    \end{tabular}
\end{table*}

\begin{figure}[htbp]
    \centering
    \includegraphics[width=0.45\textwidth]{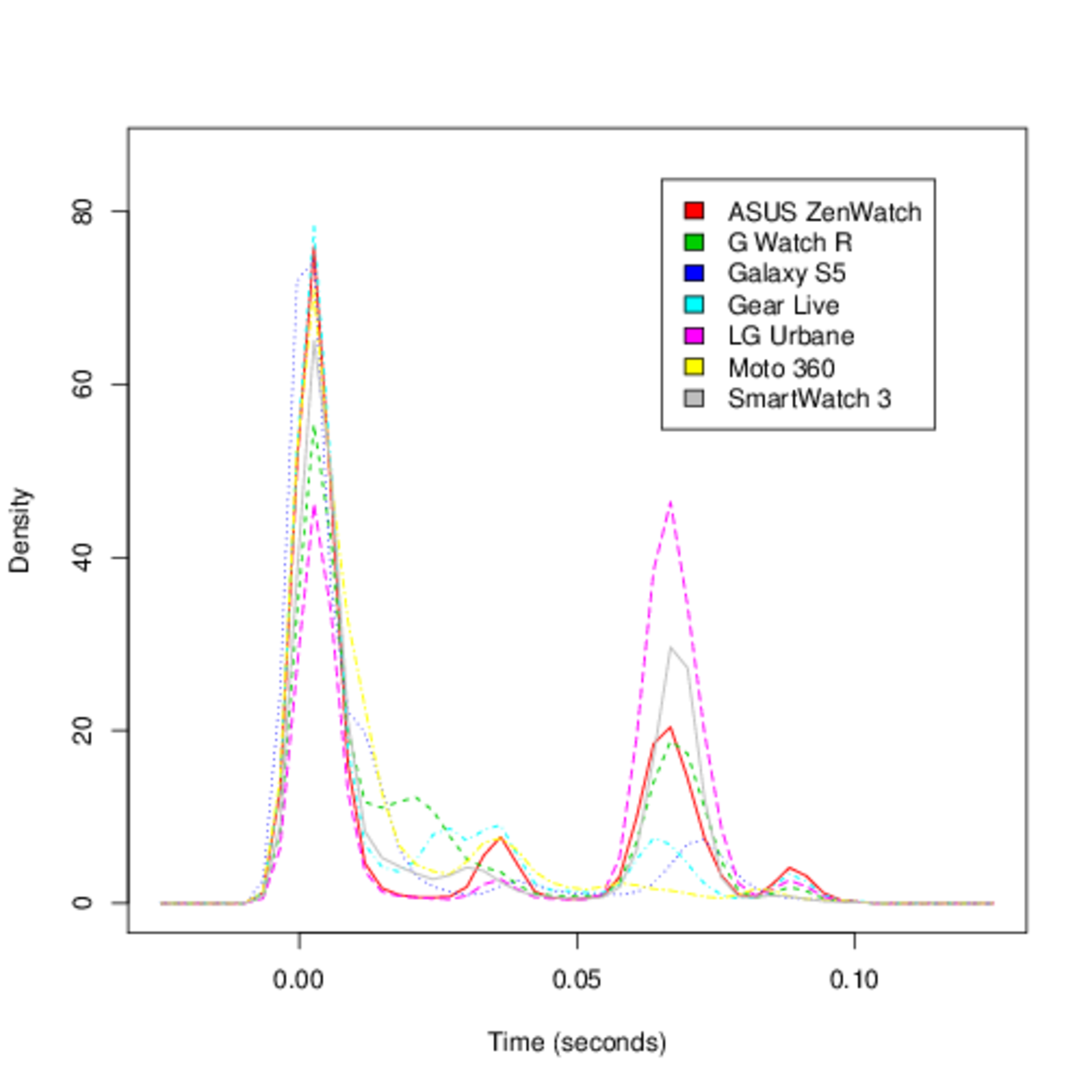}
    \caption{Inter-Arrival-Time density distributions of wearables when no packet filtering applied (i.e., no length or protocol type considered).}
    \label{fig:classic-all}
\end{figure}

Figure~\ref{fig:classic-all} displays IAT density distribution for six smartwatches and one smartphone. Although the smartphone is not a wearable device, we included here to see the impact of device diversity. In the figure, the X-axis represents packet inter-arrival-time (IAT) in seconds while the Y-axis refers to density for a given IAT value. The trend of Y-value is more important then its value. Since it is a density plot, the total area under each curve is equal to one.  As seen in the figure, for each tested wearable, there are distinct IAT distributions.  
To further understand the degree of diversity, we looked at the hardware specs for each tested wearable device and tabulated them in Table~\ref{table:classicdevicedetails}. 
 As the table suggests, in spite of different appearances and vendors, devices have small variations in their architectural details. In fact, the hardware specs correlate with density trends we observed in our tests. For instance, identical specs of ZenWatch-G, Watch R, and Sony smart watch resulted in close density plots while distinct specs of Moto 360 E9A4 and LG Urbane CFA0 resulted in further separated plots from each other.

\begin{figure*}[hbp]
         \subfloat[$pkt~length>10$ bytes]{
                 \centering
                 \includegraphics[width=0.3\textwidth]{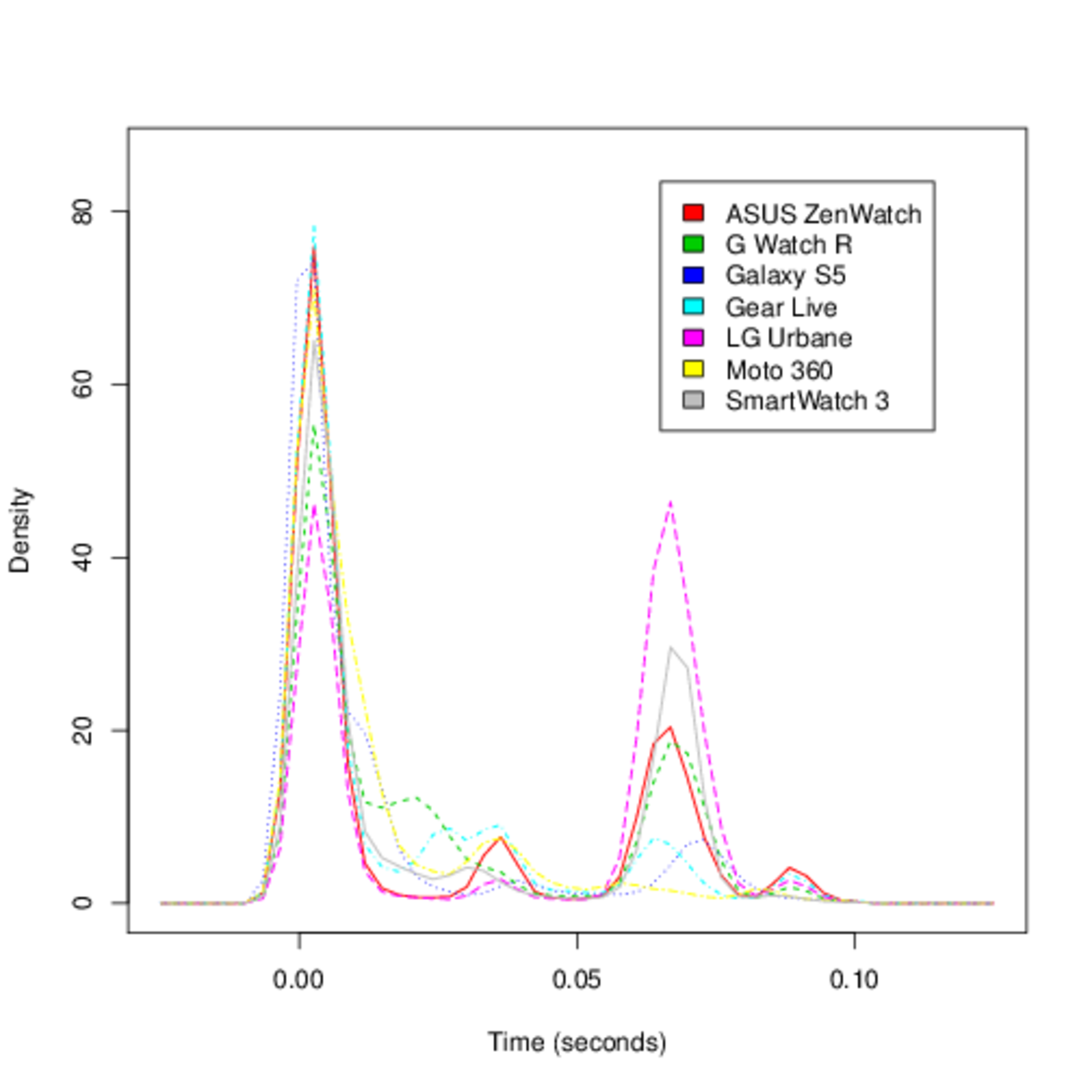}
                  \label{fig2:classic-allbylength10}
         }
          \subfloat[$pkt~length>600$ bytes]{
             \centering
             \includegraphics[width=0.3\textwidth]{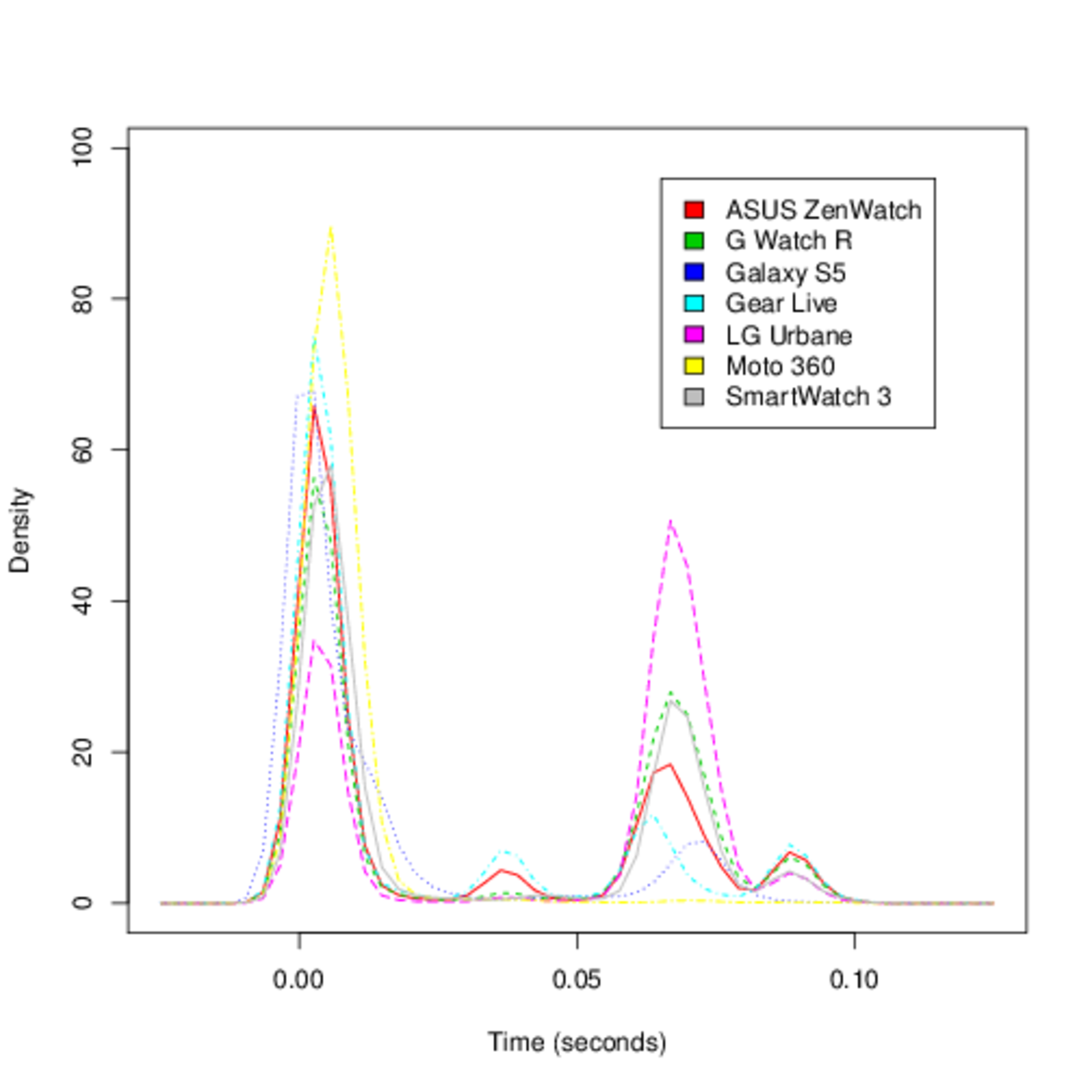}
              \label{fig2:classic-allbylength600}
      }
         \subfloat[$pkt~length>1000$ bytes]{
             \centering
             \includegraphics[width=0.3\textwidth]{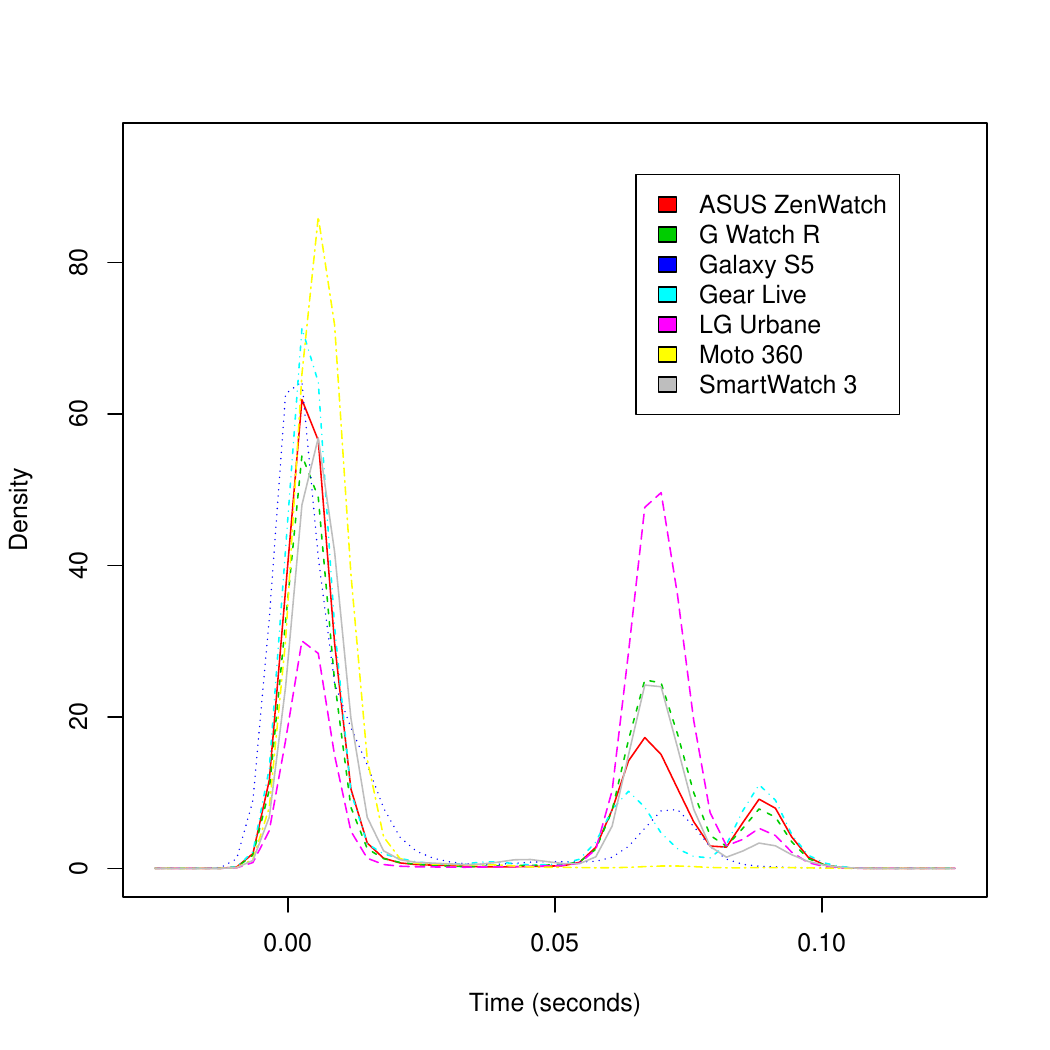}
              \label{fig2:classic-allbylength1000}
       }
 \caption{IAT density distributions of wearables with different packet lengths.} 
\label{fig:classic-allbylength}  
\end{figure*}
  
We then analyzed the effect of packet length in our wearable fingerprinting framework. Packet length means the size of evaluated packets and $length>k$ states only packets larger than $k$~bytes were considered. Figure~\ref{fig:classic-allbylength} shows the density distributions when different packet sizes were utilized. As seen in the figures, the size of the packet allows different densities for each tested wearable device.  

\begin{figure*}[htbp]
          \subfloat[$protocol==L2CAP$]{
                 \centering
                 \includegraphics[width=0.33\textwidth]{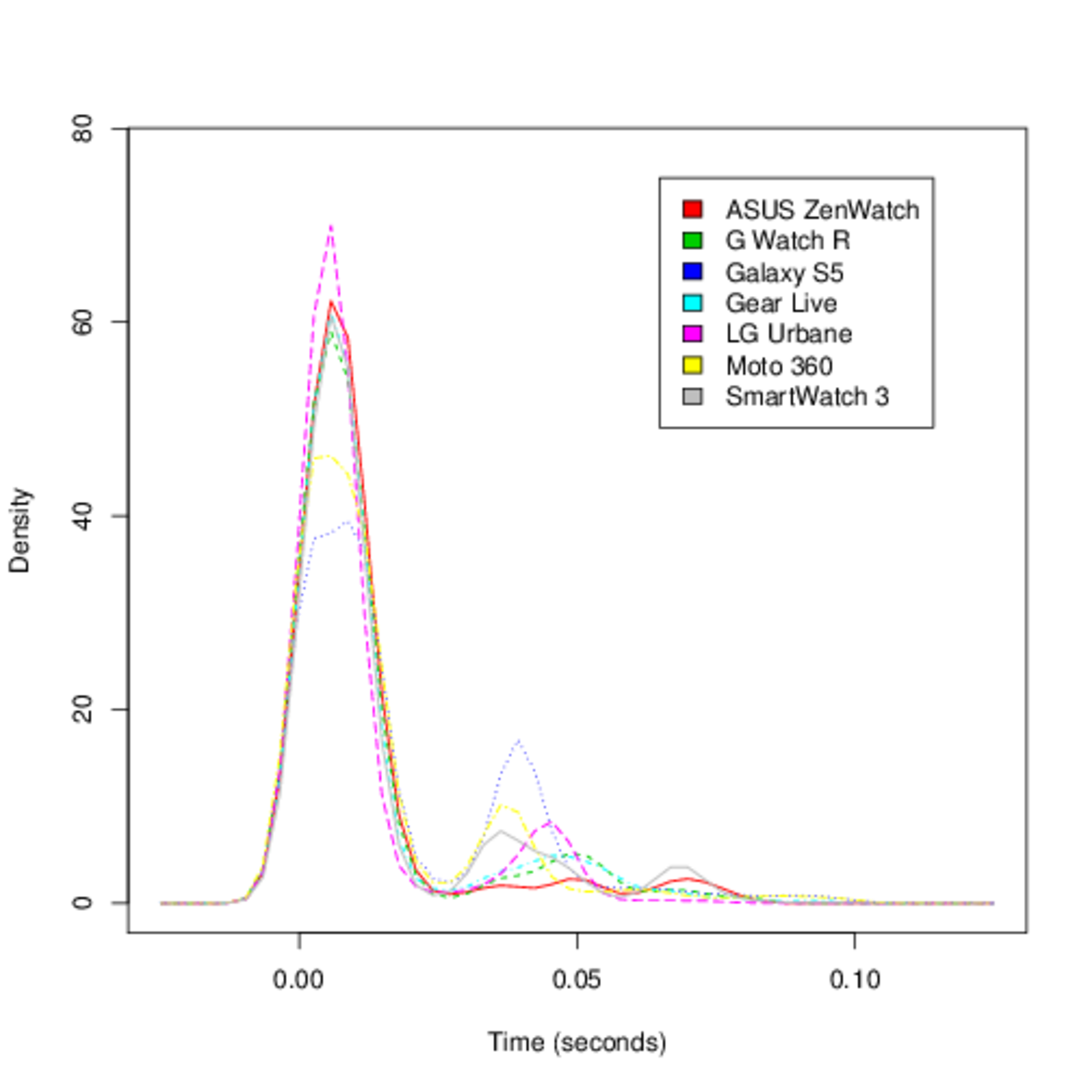}
                  \label{fig2:bossycat}
         }
         \subfloat[$protocol==RFCOMM$]{
                 \centering
                 \includegraphics[width=0.33\textwidth]{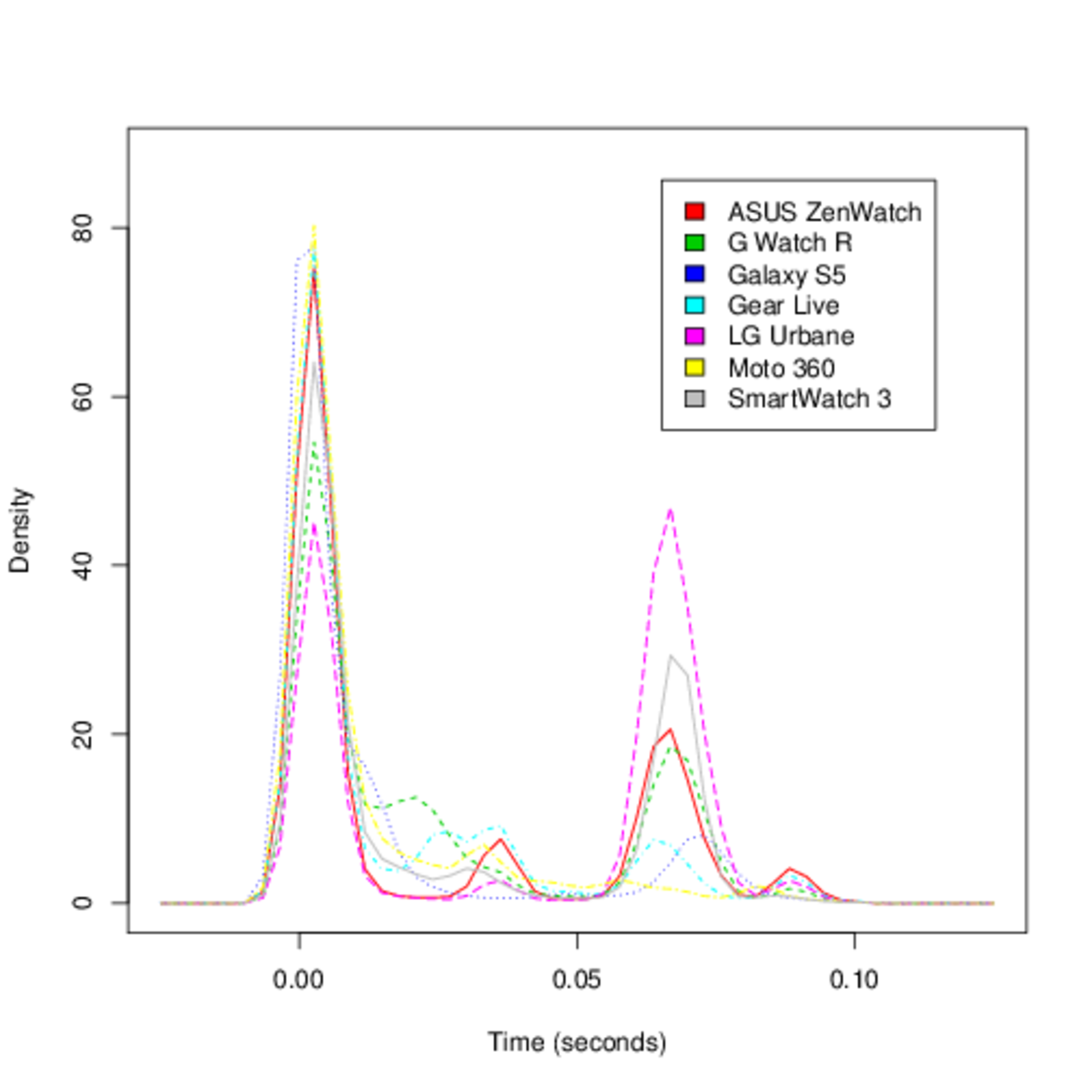}
                  \label{fig2:frowningcat}
         }
         \subfloat[$protocol==SDP$]{
             \centering
             \includegraphics[width=0.33\textwidth]{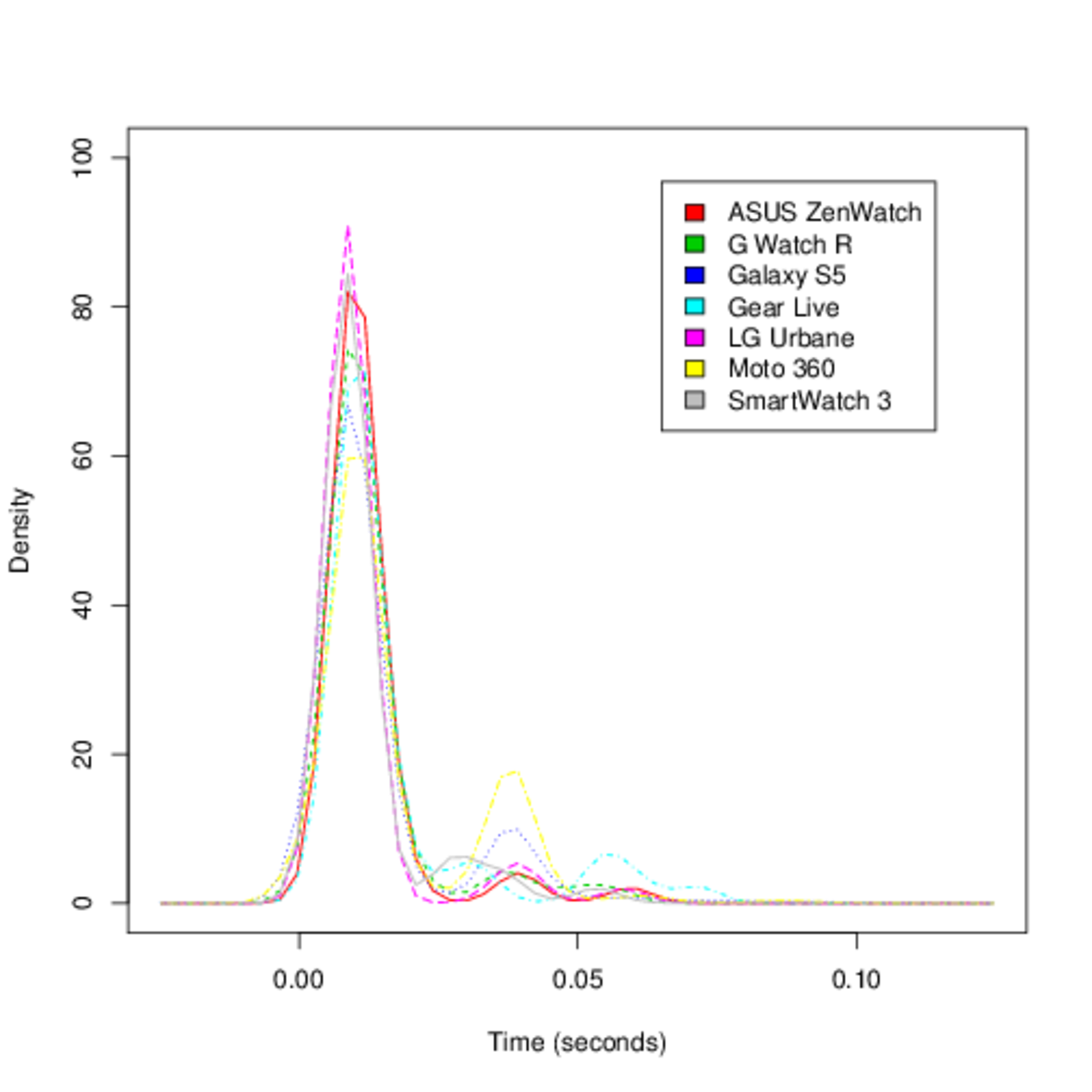}
              \label{fig2:scared}
         }
 \caption{IAT density distributions of wearables with different underlying protocol types. Density is measured in 1/seconds. Due to space limitations L2CAP, RFCOMM, and SDP are presented.} \label{fig:classic-allbyprotocol} 
\end{figure*}

\begin{figure*}[htb]
 \captionsetup{justification=centering,
  singlelinecheck=on}
          \subfloat[$protocol==HCI\_ACL$ and $pkt~length>600$]{
                 \centering
                 \includegraphics[width=0.3\textwidth]{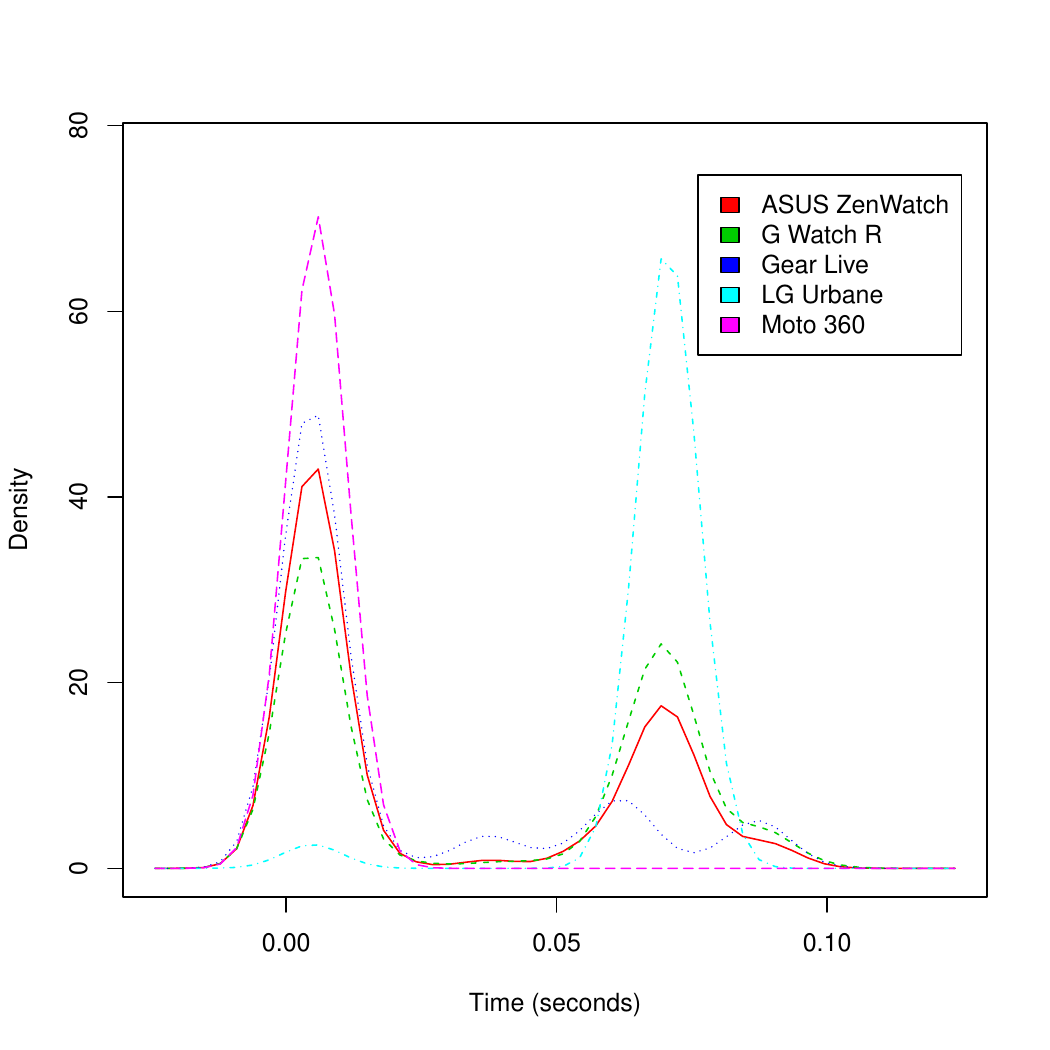}
                  \label{fig:classic-p-l-hvi-600}
         }
          \subfloat[$protocol==RFCOMM$ and $pkt~length>1000$]{
             \centering
             \includegraphics[width=0.3\textwidth]{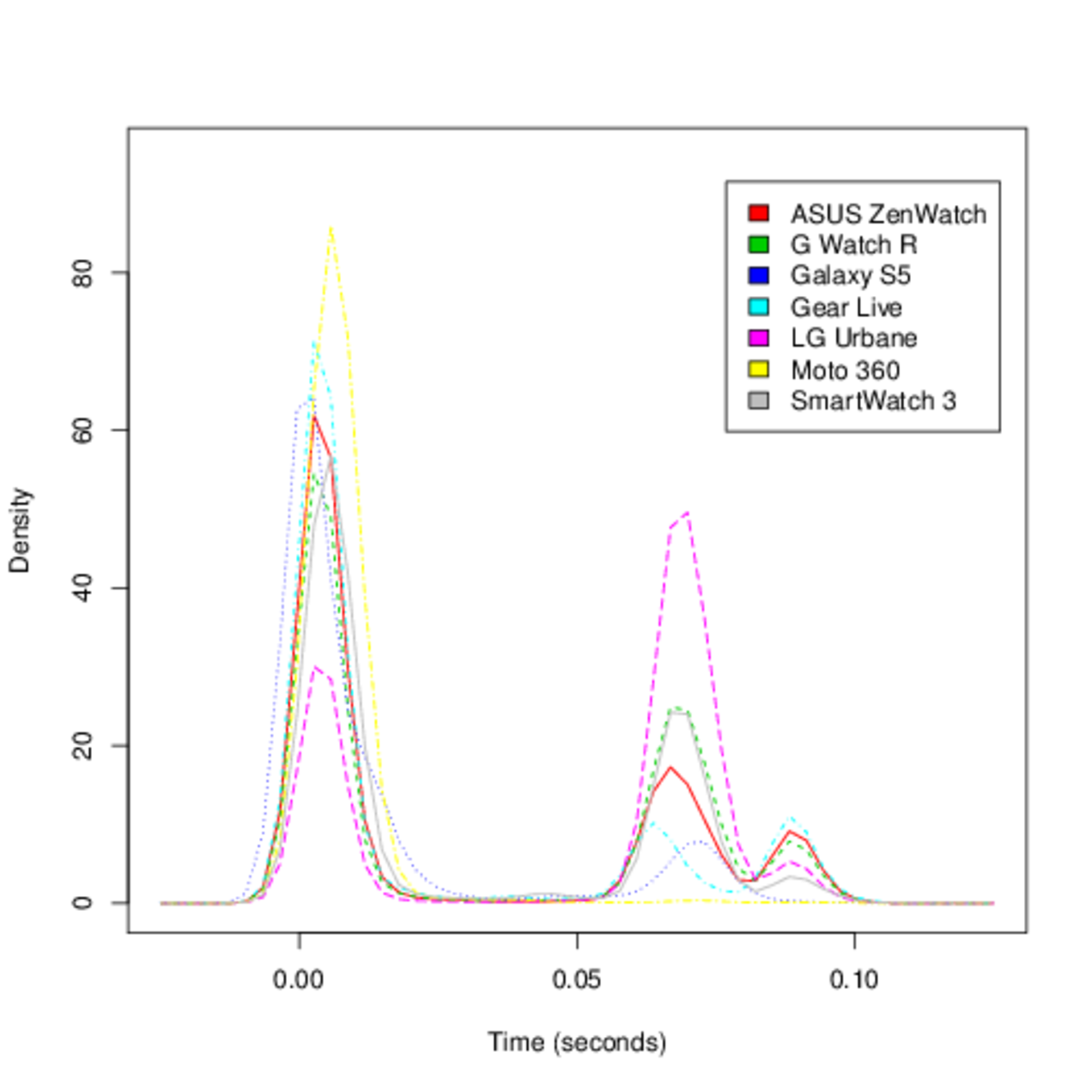}
                  \label{fig:classic-p-l-rfcomm-1000}
      }
         \subfloat[$protocol==L2CAP$ and $pkt~length>10$]{
             \centering
             \includegraphics[width=0.3\textwidth]{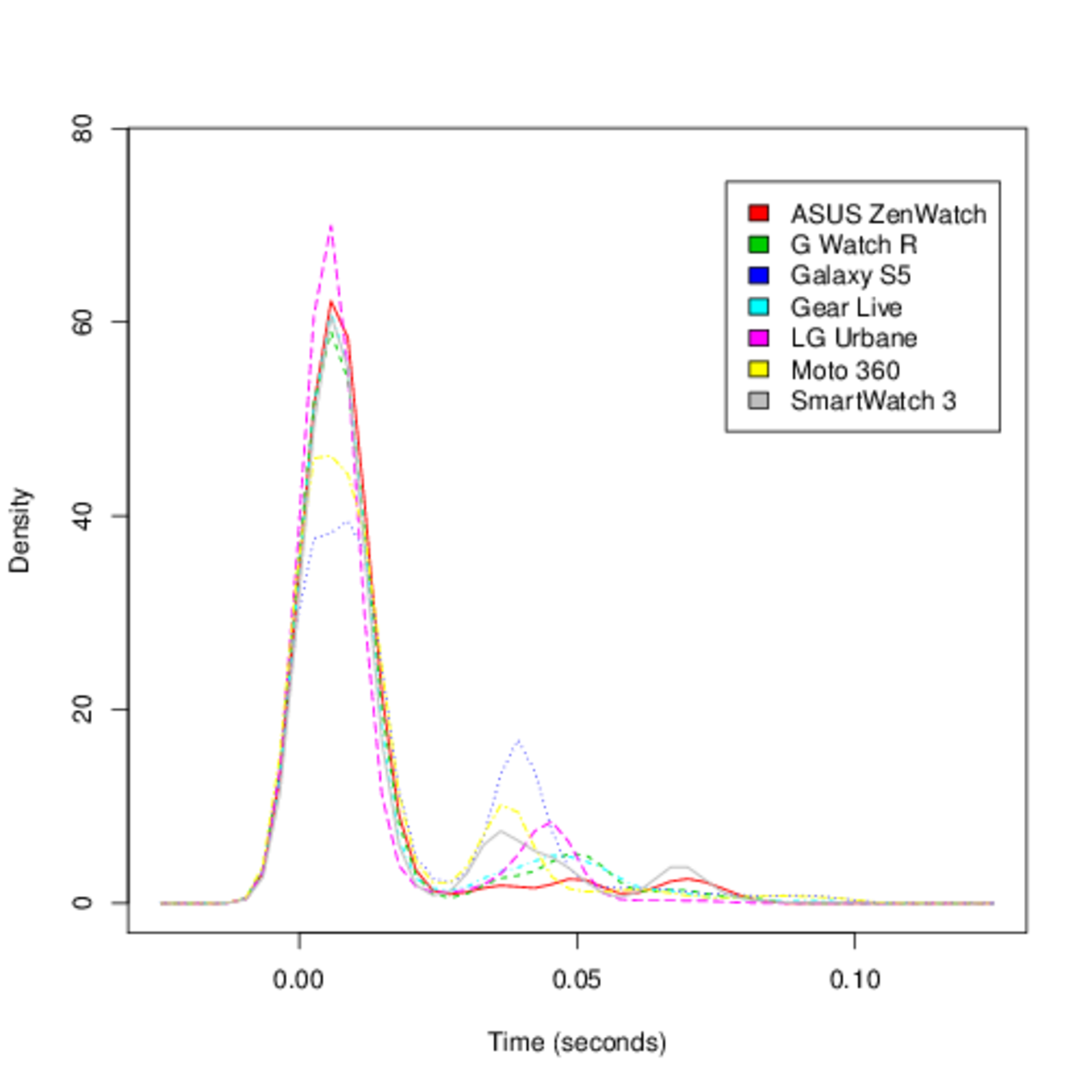}
                  \label{fig:classic-p-l-l2cap-10}
      }
       
 \caption{IAT density distributions when both packet length and underlying Bluetooth protocol type considered. Due to space limitations HCI\_ACL, L2CAP, and RFCOMM are presented.} \label{fig:classic-allbylengthandprotocol}
\end{figure*}

Next, we focused on the type of the underlying protocol utilized by the wearable device in their Bluetooth classic stack. As presented in Figure~\ref{fig:classic-allbyprotocol},
\textit{RFCOMM} provides more distinctive curves than \textit{SDP} and \textit{L2CAP} protocols. Similar to results observed with the packet length analysis, the wearable devices have varying IAT densities for each tested device. Finally, we considered both the packet length and protocol type  together. The results for this combined test case are given in Figure~\ref{fig:classic-allbylengthandprotocol}. As seen in the figures, a combined case also yields distinguishable density plots.

The density distribution curves inspired us to further analyze the feasibility of IAT feature with packet type and length in our wearable device identification framework. For this, we incorporated different machine learning algorithms into our wearable fingerprinting framework on top of the IAT feature. We divided probability density curves into 300 bins and convert the area inside each bin into a feature as described in Algorithm~\ref{alg:gensig}. Thus, each session in captures was enrolled as a signature in the database. Further, as described earlier, we used the Weka~\cite{Hall:2009:WDM:1656274.1656278} software. In addition to the ML algorithms provided in Weka, we also included an external neural network implementation with a plugin for Weka~\cite{web:WekaNeuralNetwork}. We followed Weka conventions and used 66 percent of captured sessions for learning and used the remaining for the testing phase.
Note that our framework is able to choose the best ML algorithm from the training data as explained in Section~\ref{sec:SystemModel} and Algorithm~\ref{alg:pickbest}. The framework picked \textit{Random Forest} algorithm as the best ML algorithm in the experiment.  
Table~\ref{table:classictop10} lists top-10 classification results with different filters from fingerprinting of wearables. 
$all-all$ case, in which all protocols and all packet sizes are covered (i.e., no packet length and protocol type filter is applied), provides 97\% accuracy. However, the highest accuracy, 98\%, is obtained for $RFCOMM-10$ case, where $protocol==RFCOMM$ and $pkt length > 10$ filters were applied. 
As seen in the table, our framework yields accuracy performance from 94\% to 98\% for different studied filterings and the picked ML algorithm. Thus, experiment results support that the approach to pick the most accurate ML algorithm at the training phase provides highly accurate results in the testing phase also.
 Table~\ref{table:classicaccuracy} displays the accuracy details for $RFCOMM-10$ case. Average false positive (FP) rate is lower then 1\% and both Precision and Recall are as high as 98\%. 
The proposed framework is a complementary security mechanism that is non-intrusive. In other words, it does not require running anything at the wearable device. Bluetooth traffic can be captured by the connected network device or by a third device. Traffic capturing is a passive task that does not introduce any delay and usability of wearables is not affected by the proposed framework. 
In summary, our detailed analysis and results with high accuracy and recall rates demonstrate the efficacy of our proposed wearable device identification framework.

\begin{table}[htb]
  \centering
    \caption{Top-10 classification results with different filters from fingerprinting of wearables with Bluetooth classic.} 
    \begin{tabular}{  l  l } 
        \textbf{Filtering Case} & \textbf{Accuracy (percent) }\\ \hline \hline
        RFCOMM - 10  & 98.3193  \\
        all-all  & 97.479 \\
        RFCOMM-all  & 97.479 \\
        all-600  & 96.6387 \\
        all-200  & 95.7983 \\
        RFCOMM-200  & 95.7983 \\
        all-400  & 94.958 \\
        all-800  & 94.1176 \\
        RFCOMM-600  & 94.1176 \\
        RFCOMM-400  & 94.1176 \\
     \hline
    \end{tabular}
    \label{table:classictop10}
\end{table}

 \begin{table*}[htbp]
 \scriptsize
  \centering
    \caption{Detailed accuracy results of the wearable fingerprinting framework for the identification of the wearables. }
    \begin{tabular}{ l  l l l l l l l l  } 
   \textbf{TP Rate} & \textbf{FP Rate} & \textbf{Precision} & \textbf{Recall} & \textbf{F-Measure} & \textbf{MCC}  & \textbf{ROC Area} & \textbf{PRC Area} & \textbf{Class}\\ \hline \hline
   1.000  &   0.020  &   0.913  &     1.000  &   0.955  &     0.946  &   1.000  &    1.000  &    ASUS ZenWatch\\ 
    1.000  &   0.000  &   1.000  &     1.000  &   1.000  &     1.000  &   1.000  &    1.000  &    G Watch R\\ 
    1.000  &   0.000  &   1.000  &     1.000  &   1.000  &     1.000  &   1.000  &    1.000  &    Galaxy S5\\ 
    0.947  &   0.000  &   1.000  &     0.947  &   0.973  &     0.968  &   1.000  &    1.000  &    Gear Live\\ 
    0.941  &   0.000  &   1.000  &     0.941  &   0.970  &     0.965  &   0.999  &    0.991  &    LG Urbane\\ 
    1.000  &   0.000  &   1.000  &     1.000  &   1.000  &     1.000  &   1.000  &    1.000  &    Moto 360\\ 
   1.000  &   0.000  &   1.000  &     1.000  &   1.000  &     1.000  &   1.000  &    1.000  &    SmartWatch 3\\ \hline \hline
    0.983  &   0.004  &   0.985  &     0.983  &   0.983  &     0.980  &   1.000  &    0.999  & Weighted Avg. \\   
     \hline
    \end{tabular}
    \label{table:classicaccuracy}
\end{table*}

\section{Security Impact: Threats \& Use Cases }\label{sec:securityImpact}  
\subsection{Threats}
In a wearable networking environment, the network can dynamically grow and shrink in size with new wearable devices and equipment depending on the usage. 
New wearables can join and leave the network and device configurations can change dynamically more frequently than usual networks. This situation, unfortunately, poses challenges to the security posture of a network.  
Specifically, adversaries may target the functions of the wearable devices or a network with wearables as follows:  
\begin{itemize}
\item \textit{(1) Unauthorized wearables with correct credentials}: A network with wearables may include unauthorized devices with legitimate credentials. For instance, wearable devices could be authenticated to the network via an authorized user of the authentication realm for a specific purpose but could still be part of the network even beyond their intended duration. 

\item \textit{(2) Wearable devices with counterfeit components}: In a network with wearables, there may be legitimate wearables devices with counterfeit architectural (internal) components (e.g., memory, chip)~\cite{DODJan13Report, web:Stecklow}.

\item \textit{(3) Outsider wearable devices (brute-force attackers)}: A network with wearables may include an outsider, whose primary focus is to attempt to participate in the wearables network by exhaustively searching for the correct credentials. 

\item \textit{(4) Information-leaking wearables}: A network with wearables may include an active outsider or compromised insider device that tries to leak important information about the network. 
\end{itemize}

\begin{figure}[H]
    \centering
    \includegraphics[clip, trim=0.5cm 0cm 0.5cm 0cm, width=0.55\textwidth]{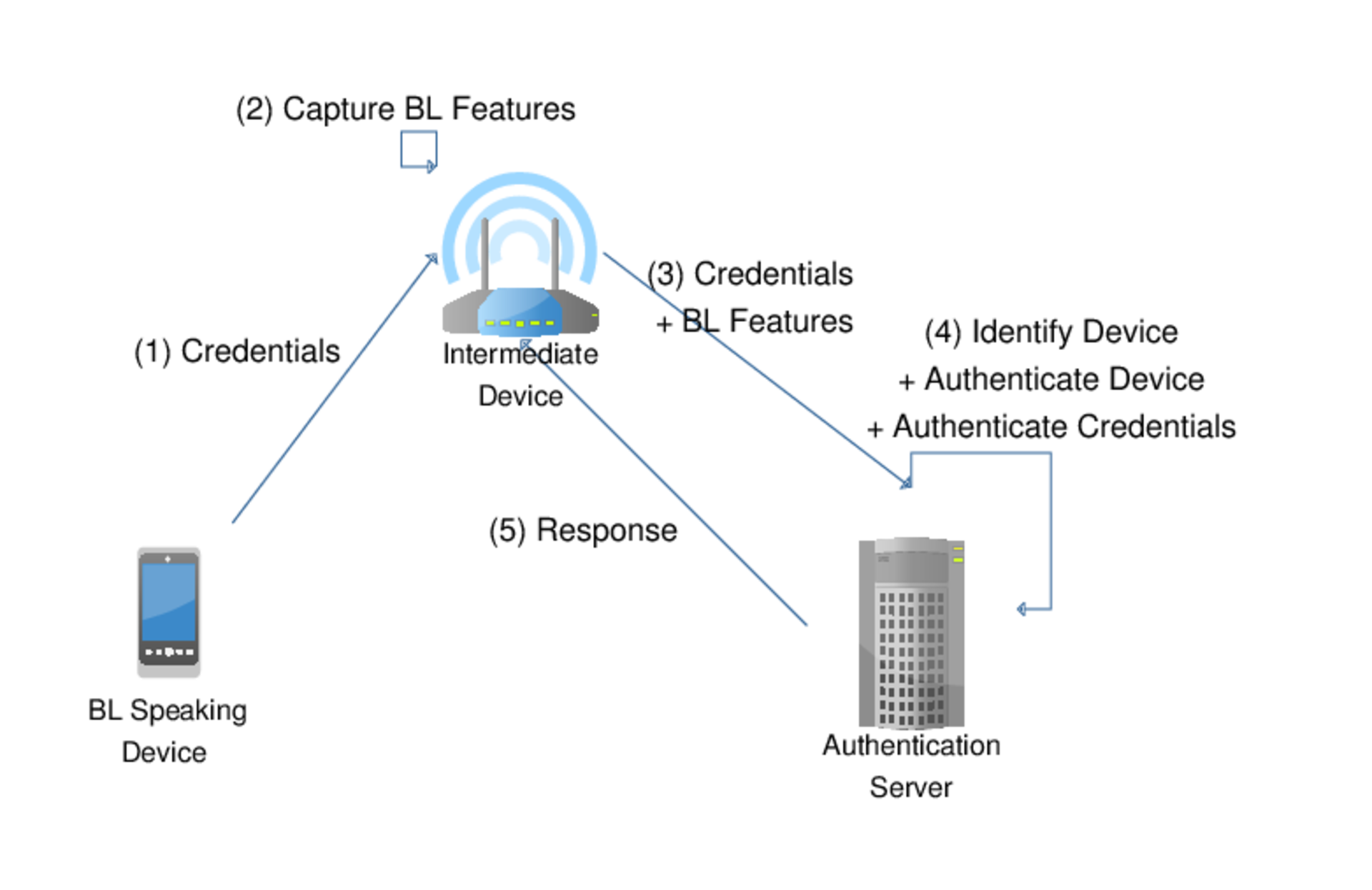}
    \caption{A possible scenario where the proposed framework is utilized by an authentication server to check device identity in addition to user credentials.}
    \label{fig:depsystem}
\end{figure}

\subsection{Use Cases for Security}   

Current security models were mostly built to verify the user, not the device itself. However, a non-intrusive wearable fingerprinting mechanism such as the one proposed in this paper can 
complement existing security mechanisms against these threats. Figure~\ref{fig:depsystem} illustrates a sample use of the proposed fingerprinting technique with a traditional authentication system. When a  connection request from a BL speaking device with (1) credentials is received by intermediate device (e.g., smart phone, laptop), this intermediate device will (2) capture BL features and then (3) forward it with credentials to the authentication server. In addition to traditional credential controls, (4) authentication server will use fingerprinting techniques to identify the device and control whether this device should be allowed to connect the network with given credentials. Enabling such additional controls would enhance the authentication server's ability to address the aforementioned threats in the following ways:   
\begin{itemize}
\item  (1) The wearable fingerprinting technique can detect unauthorized wearables with correct credentials as authentication server would control whether the device is authorized to access the network in addition to credentials control. Furthermore, the fingerprinting technique can detect unmanageable wearables with mechanisms along with traditional access control mechanisms (e.g., Network Access Control (NAC)), improving the efficacy of such products.
 \item  (2) It can be used to help detect wearable devices with counterfeit or corrupt components as fingerprints depend on both software and hardware components of devices. As a device with counterfeit or corrupt components will display different performance metrics (e.g., packet production rate) when compared with genuine one, proposed fingerprinting technique would not identify such devices as authentic ones.
 \item  (3) It can be helpful in determining outsider wearable devices that use high resources to perform brute-force attacks. As wearable devices are typically resource-limited, outsiders can use rogue wearables to perform attacks inside the network. Such rogue wearables will have different performance metrics which can be detected by the proposed fingerprinting technique. 
\item  (4) It can be used to identify resource-limited wearable devices accessing the network and further enforce access control targeted for such devices. A smartwatch like wearable device can be used to leak information from a network where mobility of physical devices like laptops, usb devices are restricted. An insider may attempt to use a wearable device like a smartwatch (some of which looks identical to regular watches) to access a restricted network and leak information to outside. In addition to credentials control, detection of accesses from such wearable devices helps to address such threats. 
 \end{itemize}
 
Thus, the machine learning based fingerprinting framework proposed in this work infers the device type information as a cyber intelligence. This intelligence can be utilized by any authentication or authorization function in the network where both the claimed identity of a device and any attempted action can be further checked with the actual device type to determine whether this is a suspicious activity, i.e., a cyber threat. 
Obviously, such cyber threat intelligence would increase the overall security posture of the network. 

\section{Conclusion}\label{sec:Conclusion}
Cyberspace is expanding quickly with the introduction of new wearable devices (e.g., smart watches). Given the increasingly critical nature of the cyberspace of these wearable devices, it is imperative that they are secured. An adversary only needs one entry point to infiltrate networks. Nonetheless, the current security mechanisms are focused on validating the user, not the device itself. An unauthorized wearable device even with an authorized user can perpetrate malicious activities. Hence, in this work, we considered wearable fingerprinting as a non-intrusive complementary security mechanism for wearables. Specifically, we introduced a wearable fingerprinting framework focusing on the characteristics of Bluetooth classic protocol, which is a common protocol used in the wearables realm.  Our framework also included a comprehensive set of Machine Learning (ML) algorithms (20 different ML algorithm) in the classification process to pick the best performing algorithm. Furthermore, we evaluated the performance of our wearable fingerprinting technique on real wearable devices. Our evaluation demonstrated the functionality and feasibility of the proposed technique. Specifically, our detailed accuracy results show on average 98.5\%, 98.3\% precision and recall for wearables using the Bluetooth classic protocol. 
In essence, the proposed machine learning based fingerprinting framework provides reliable device type information to any authentication or authorization point in the network where the claimed identity or attempted action can be further checked to determine whether this is a suspicious activity, i.e., a cyber threat. Certainly, such cyber threat intelligence would improve the overall security posture of emerging IoT networks with multiple wearable devices.

\ifCLASSOPTIONcompsoc
  \section*{Acknowledgments}
\else
  \section*{Acknowledgment}
\fi

Authors acknowledge US Air Force Research Lab's (AFRL-FA8750-13-2-0116) and National Science Foundation's (NSF-CAREER-CNS-1453647)  support in this work. Any opinions, findings and conclusions or recommendations expressed in this material are those of the authors and do not necessarily reflect the views of the funding agencies.

\ifCLASSOPTIONcaptionsoff
  \newpage
\fi


\begin{thebibliography}{10} \scriptsize
\providecommand{\url}[1]{#1}
\csname url@samestyle\endcsname
\providecommand{\newblock}{\relax}
\providecommand{\bibinfo}[2]{#2}
\providecommand{\BIBentrySTDinterwordspacing}{\spaceskip=0pt\relax}
\providecommand{\BIBentryALTinterwordstretchfactor}{4}
\providecommand{\BIBentryALTinterwordspacing}{\spaceskip=\fontdimen2\font plus
\BIBentryALTinterwordstretchfactor\fontdimen3\font minus
  \fontdimen4\font\relax}
\providecommand{\BIBforeignlanguage}[2]{{%
\expandafter\ifx\csname l@#1\endcsname\relax
\typeout{** WARNING: IEEEtran.bst: No hyphenation pattern has been}%
\typeout{** loaded for the language `#1'. Using the pattern for}%
\typeout{** the default language instead.}%
\else
\language=\csname l@#1\endcsname
\fi
#2}}
\providecommand{\BIBdecl}{\relax}
\BIBdecl

\bibitem{web:Kelly}
S.~M. Kelly, ``Experts: Internet of things and wearables will dominate by
  2025,'' July 2015, \url{http://mashable.com/2014/05/14/pew-iot-study}.

\bibitem{IoT_Evolution}
\BIBentryALTinterwordspacing
D.~Evans, ``The {I}nternet of {T}hings: How the next evolution of the
  {I}nternet is changing everything,'' Apr. 2011. [Online]. Available:
  \url{https://www.cisco.com/web/about/ac79/docs/innov/IoT_IBSG_0411FINAL.pdf}
\BIBentrySTDinterwordspacing

\bibitem{IoT_Europe}
\BIBentryALTinterwordspacing
P.~F. Drucker, ``{Internet of Things} position paper on standardization for
  {IoT} technologies,'' Jan. 2015. [Online]. Available:
  \url{http://www.internet-of-things-research.eu/pdf/IERC_Position_Paper_IoT_Standardization_Final.pdf}
\BIBentrySTDinterwordspacing

\bibitem{watch}
J.~Comstock,
  ``{http://mobihealthnews.com/37543/pwc-1-in-5-americans-owns-a-wearable-1-in-10-wears-them-daily/},''
  \emph{Mobi Health News}, 2014.

\bibitem{miller2005coping}
R.~E. Miller, D.-L. Chen, D.~Lee, and R.~Hao, ``Coping with nondeterminism in
  network protocol testing,'' in \emph{Testing of Communicating Systems}.\hskip
  1em plus 0.5em minus 0.4em\relax Springer, 2005, pp. 129--145.

\bibitem{caballero2007fig}
J.~Caballero, S.~Venkataraman, P.~Poosankam, M.~G. Kang, D.~Song, and A.~Blum,
  ``Fig: Automatic fingerprint generation,'' \emph{Department of Electrical and
  Computing Engineering}, p.~27, 2007.

\bibitem{DODJan13Report}
\texttt{Office of The Secretary of The Department of Defense}, ``Resilient
  military systems and the advanced cyber threat final report,'' in
  \emph{Defense Science Board Task Force on Resilient Military Systems}, Jan
  2013.

\bibitem{web:Stecklow}
S.~Stecklow, ``Exclusive: U.s. nuclear lab removes chinese tech over security
  fears,'' Jan 2013,
  \url{http://www.reuters.com/article/2013/01/07/us-huawei-alamos-idUSBRE90608B20130107}.

\bibitem{Wenyi-Maca-InfocomWS-14}
W.~Liu, A.~Uluagac, and R.~Beyah, ``Maca: A privacy-preserving multi-factor
  cloud authentication system utilizing big data,'' in \emph{Computer
  Communications Workshops (INFOCOM WKSHPS), 2014 IEEE Conference on}, April
  2014, pp. 518--523.

\bibitem{1059392}
T.~Kohno, A.~Broido, and K.~Claffy, ``Remote physical device fingerprinting,''
  in \emph{2005 IEEE Symposium on Security and Privacy (S P'05)}, May 2005, pp.
  211--225.

\bibitem{1409958}
S.~Jana and S.~K. Kasera, ``On fast and accurate detection of unauthorized
  wireless access points using clock skews,'' \emph{IEEE Transactions on Mobile
  Computing}, vol.~9, no.~3, pp. 449--462, March 2010.

\bibitem{WDF:5a}
L.~Letaw, J.~Pletcher, and K.~Butler, ``Host identification via usb
  fingerprinting,'' \emph{Systematic Approaches to Digital Forensic Engineering
  (SADFE)}, 2011.

\bibitem{Danev:2012:PIW:2379776.2379782}
\BIBentryALTinterwordspacing
B.~Danev, D.~Zanetti, and S.~Capkun, ``On physical-layer identification of
  wireless devices,'' \emph{ACM Comput. Surv.}, vol.~45, no.~1, pp. 6:1--6:29,
  Dec. 2012. [Online]. Available:
  \url{http://doi.acm.org/10.1145/2379776.2379782}
\BIBentrySTDinterwordspacing

\bibitem{bluetooth}
J.~Hall, M.~Barbeau, and E.~Kranakis, ``Rogue devices in bluetooth networks
  using radio frequency fingerprinting,'' in \emph{IASTED International Conf.
  on Communications and Computer Networks (CCN)}, 2006.

\bibitem{wifi}
V.~Brik, S.~Banerjee, M.~Gruteser, and S.~Oh, ``Wireless device identification
  with radiometric signatures,'' in \emph{Proc. of the 14th ACM International
  Conf. on Mobile Computing and Networking (MobiCom)}, 2008.

\bibitem{Uluagac-GTID-CNS}
S.~Uluagac, S.~V. Radhakrishnan, C.~L. Corbett, A.~Baca, and R.~Beyah, ``A
  passive technique for fingerprinting wireless devices with wired-side
  observations,'' in \emph{2013 IEEE Conference on Communications and Network
  Security (CNS) (IEEE CNS 2013)}, Washington, USA, Oct. 2013, pp. 471--479.

\bibitem{Sakthi-GTID-TDSC}
S.~V. Radhakrishnan, A.~S. Uluagac, and R.~Beyah, ``Gtid: A technique for
  physical device and device type fingerprinting,'' \emph{IEEE Transactions on
  Dependable and Secure Computing}, vol.~99, no. PrePrints, p.~1, 2015.

\bibitem{7239531}
Q.~Xu, R.~Zheng, W.~Saad, and Z.~Han, ``Device fingerprinting in wireless
  networks: Challenges and opportunities,'' \emph{Communications Surveys
  Tutorials, IEEE}, vol.~18, no.~1, pp. 94--104, Firstquarter 2016.

\bibitem{web:Tizen}
``Tizen operating system,'' 2015, \url{https://www.tizen.org}.

\bibitem{web:pebble}
``Pebble operating system,'' 2015, \url{https://blog.getpebble.com}.

\bibitem{web:bluetoothdate}
``The story behind bluetooth technology,'' 2015,
  \url{https://www.bluetooth.com/what-is-bluetooth-technology/bluetooth}.

\bibitem{web:bluetoothsig}
``History of bluetooth,'' 2015,
  \url{https://www.bluetooth.com/media/our-history}.

\bibitem{5686874}
G.~Shu and D.~Lee, ``A formal methodology for network protocol
  fingerprinting,'' \emph{Parallel and Distributed Systems, IEEE Transactions
  on}, vol.~22, no.~11, pp. 1813--1825, Nov 2011.

\bibitem{web:Scapy}
``Python scapy,'' 2015, \url{http://www.secdev.org/projects/scapy}.

\bibitem{web:tcpdump}
``Tcpdump: A packet analyzer tool,'' 2015, \texttt{http://www.tcpdump.org/}.

\bibitem{web:wireshark}
``Wireshark,'' 2015, \url{https://www.wireshark.org}.

\bibitem{web:ubertooth}
``Project ubertooth,'' 2015, \url{http://ubertooth.sourceforge.net}.

\bibitem{Hall:2009:WDM:1656274.1656278}
\BIBentryALTinterwordspacing
M.~Hall, E.~Frank, G.~Holmes, B.~Pfahringer, P.~Reutemann, and I.~H. Witten,
  ``The weka data mining software: An update,'' \emph{SIGKDD Explor. Newsl.},
  vol.~11, no.~1, pp. 10--18, Nov. 2009. [Online]. Available:
  \url{http://doi.acm.org/10.1145/1656274.1656278}
\BIBentrySTDinterwordspacing

\bibitem{web:WekaNeuralNetwork}
``Java (convolutional or fully-connected) neural network implementation with
  plugin for weka. uses dropout and rectified linear units.'' 2016,
  \texttt{https://github.com/amten/NeuralNetwork}.

\end{thebibliography}


 \begin{IEEEbiography}[{\includegraphics[width=1in,height=1.25in,clip,keepaspectratio]{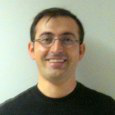}}]{Hidayet AKSU} 
 received his Ph.D., M.S. and B.S. degrees from Bilkent University, all in Department of Computer Engineering, in 2014, 2008 and 2005, respectively. He is currently a Postdoctoral Associate in the Department of Electrical \& Computer Engineering at Florida International University (FIU). Before that, he worked as an Adjunct Faculty in the Computer Engineering Department of Bilkent University. He conducted research as visiting scholar at IBM T.J. Watson Research Center, USA in 2012-2013. He also worked for Scientific and Technological Research Council of Turkey (TUBITAK).
  His research interests include security for cyber-physical systems, internet of things, security for critical infrastructure networks, IoT security, security analytics, social networks, big data analytics, distributed computingx, wireless networks, wireless ad hoc and sensor networks, localization, and p2p networks.
\end{IEEEbiography} 

\begin{IEEEbiography}[{\includegraphics[width=1in,height=1.25in,clip,keepaspectratio]{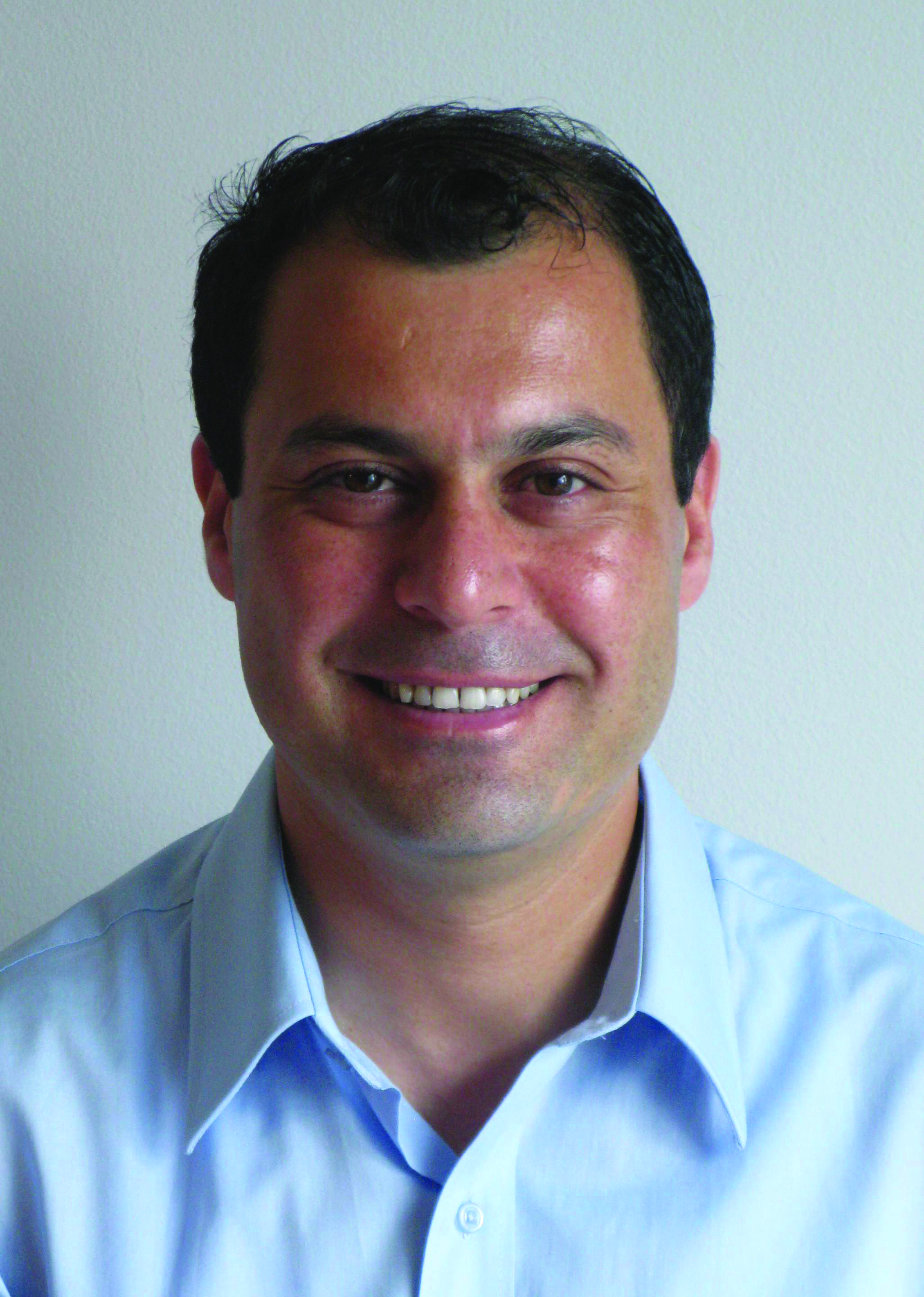}}]{Dr. A. Selcuk Uluagac} is currently an Assistant Professor in the Department of Electrical and Computer Engineering (ECE) at Florida International University (FIU). Before joining FIU, he was a Senior Research Engineer in the School of Electrical and Computer Engineering (ECE) at Georgia Institute of Technology. Prior to Georgia Tech, he was a Senior Research Engineer at Symantec. He earned his Ph.D. with a concentration in information security and networking from the School of ECE, Georgia Tech in 2010. He also received an M.Sc. in Information Security from the School of Computer Science, Georgia Tech and an M.Sc. in ECE from Carnegie Mellon University in 2009 and 2002, respectively. 
The focus of his research is on cyber security topics with an emphasis on its practical and applied aspects. He is interested in and currently working on problems pertinent to the security of Cyber-Physical Systems and Internet of Things. In 2015, he received a Faculty Early Career Development (CAREER) Award from the US National Science Foundation (NSF).
In 2015, he was awarded the US Air Force Office of Sponsored Research (AFOSR)’s 2015 Summer Faculty Fellowship. 
He is also an active member of IEEE (senior grade), ACM, and ASEE and a regular contributor to national panels and leading journals and conferences in the field. Currently, he is the area editor of Elsevier Journal of Network and Computer Applications and serves on the editorial board of the IEEE Communication Surveys and Tutorials. More information can be obtained from: http://web.eng.fiu.edu/selcuk.
\end{IEEEbiography}

\begin{IEEEbiographynophoto}
{Elizabeth Serena Bentley}  has a B.S. degree in Electrical Engineering from Cornell University, a M.S. degree in Electrical Engineering from Lehigh University, and a Ph.D. degree in Electrical Engineering from University at Buffalo. She was a National Research Council Post-Doctoral Research Associate at the Air Force Research Laboratory in Rome, NY. Currently, she is employed by the Air Force Research Laboratory in Rome, NY, performing in-house research and development in the Networking Technology branch. Her research interests are in cross-layer optimization, wireless multiple-access communications, wireless video transmission, modeling and simulation, and directional antennas/directional networking.
\end{IEEEbiographynophoto}

\end{document}